\documentclass[manuscript]{emulateapj}
\usepackage{amsmath}

\shorttitle{Distortion of the luminosity function of galaxies}
\shortauthors{Fialkov \& Loeb}

\begin{document}

\title{Distortion of the luminosity function of high-redshift 
galaxies by gravitational lensing}

\author{Anastasia Fialkov}
\affil{Departement de Physique, Ecole Normale Superieure, CNRS, 24 rue Lhomond, Paris, 75005 France}
\email{anastasia.fialkov@phys.ens.fr}
\and
\author{Abraham Loeb}
\affil{Department of Astronomy, Harvard University,
60 Garden Street, $MS-51$, Cambridge, MA, 02138 U.S.A.}
\email{aloeb@cfa.harvard.edu}

\begin{abstract}

The observed properties of high redshift galaxies depend on the underlying foreground distribution of large scale structure, which distorts their intrinsic properties via gravitational lensing. We focus on the regime where the dominant contribution originates from a  single lens and examine the statistics of gravitational lensing by a population of virialized and non-virialized structures using sub-mm galaxies at $z\sim 2.6$ and Lyman-break galaxies at redshifts $z\sim 6-15$ as the background sources. We quantify the effect of lensing  on the luminosity function of the high redshift sources, focusing on  the intermediate and small magnifications, $\mu\la 2$, which affect the majority of the background galaxies,  and comparing to the case of strong lensing.  We show that, depending on the intrinsic properties of the background galaxies, gravitational lensing can significantly  affect the observed luminosity function even when no obvious strong lenses are present. Finally, we find that in the case of the Lyman-break galaxies it is important to account for the surface brightness profiles of both the foreground and the background galaxies when computing the lensing statistics, which introduces a selection criterion for the background galaxies that can actually be observed. Not taking this criterion into account leads to an overestimation of the number densities of very bright galaxies by nearly two orders of magnitude.  
\end{abstract}


\section{Introduction}

Detection of the high-redshift galaxies is a primary frontier in observational cosmology. Sampling and analyzing the properties of different types of high-redshift sources will constrain galaxy formation and star formation histories at different epochs, and explain their role in reionization and metal enrichment of the Universe \citep{LF,Ellis}.    For instance, dusty star-forming galaxies at a redshift range $z\sim 2-4$   are the most luminous galaxies at that epoch  and  host a considerable fraction of star formation at $z\geq 2$ \citep{Hughes:1998, Blain:1999, Chapman:2005}, which includes the epoch when the star-formation rate density peaked \citep{Madau:1996}.  Despite their leading role in the history of galaxy formation,  this population of galaxies has not been well studied yet due to dust obscuration, and is  surveyed at present at sub-millimeter wavelengths. A reliable number counts of resolved sources from this population are being provided for the first time  \citep{Karim:2013} by the Atacama Large Millimeter/submillimeter Array (ALMA). 
 Another population of high redshift sources which is not well constrained at the moment  are the galaxies existing during the epoch of reionization and observed  in their rest-frame UV at redshifts out to $z\sim 10$ by the Wide Field Camera 3 infrared channel (WFC3/IR, \citet{Kimble:2008}) on board  the Hubble Space Telescope (HST), with  plans  to push this frontier to even higher redshifts with the James Webb Space Telescope (JWST). The identification of these galaxies is done based on the Lyman-break technique which relies on the  absorption of ultra-violet photons at wavelengths shortward of the redshifted Ly-$\alpha$ line due to neutral hydrogen fraction. Because the UV luminosity of the Lyman-break galaxies strongly correlates with the star formation rates \citep{Madau:1998}, establishing the UV luminosity function at high redshift is an essential step towards measuring the halo abundances at these redshifts (e.g., via abundance-matching techniques \citep{Kravtsov:2004, Tasitsiomi:2004,   Vale:2004, Conroy:2006}) and assessing the impact of these galaxies on the reionization of cosmic hydrogen in the first billion years after the Big Bang. 

 The observable properties of any high-redshift population of sources, such as their abundance and luminosity function, differ from the intrinsic ones, since their radiation is subject to gravitational lensing. Along a random line of sight, photons are deflected  on their path from the distant galaxies to the observer, a process which can be described statistically \citep{Turner:1984,Pei:1993, Pei:1995, Perrotta:2002,   Negrello:2007, Lima:2010a, Lima:2010b, Jain:2011, Wardlow:2013}. In rare cases when  the minimal separation between the center of the foreground structure and the trajectory of the photon (referred  to as its impact parameter)  is small enough, gravitational lensing has a strong effect on the light rays leading to a significant magnification of the background galaxy, multiple images and a major distortion of the source appearance on the sky.  The photon deflection has two main impacts on the observed sources: the fluxes are magnified  and the solid angle within which the sources are observed is increased. Both effects need to be taken into account when inferring the intrinsic properties of a background population from the observed ones. The efficiency of lensing depends on the properties of both the foreground and the background populations. For instance, for the same set of lenses the main contribution to strong lensing with magnification, $\mu$,  above 2 originates from structures at a mean redshift of $\bar z\sim 0.6$ when the sources are the sub-millimeter galaxies (SMGs) located at $z\sim 2$; while the Lyman-break galaxies (LBGs) at $z\sim 6-10$ are most affected by foreground objects at $\bar z\sim 1$. However, for any background population the strongest effect of lensing is expected to be on the rare brightest sources. This is because an observed bright source has an enhanced likelihood of being a magnified intrinsically fainter galaxy, which are much more abundant, thus leading to an overestimation of the intrinsically bright population and distorted appearance of its luminosity function.

Lensing appears to be a useful tool when counting the numbers of high redshift galaxies and measuring their luminosities. The magnification makes the selection of strongly lensed galaxies easy for large area galaxy surveys. For instance, more than $85\%$ of dusty sub-millimeter  galaxies observed by the {\it Herschel} Space Observatory were confirmed to be gravitationally lensed by an intervening foreground structure  along the line of sight \citep{Negrello:2010,  Wardlow:2013,  Bussmann:2013}. In addition, faint galaxies behind massive foreground structures can be magnified above the detection limit.   When background samples behind known foreground lens distribution are considered, the distortions introduced by strong gravitational lensing can be easily spotted and corrected for.  For example,  in the framework of the Hubble Frontier Fields program  faint background galaxies at $z\sim 7$ magnified by up to a factor of $\sim30 $ by foreground massive galaxy clusters  were detected down to the absolute magnitude of $M_{UV}\sim -15.5$ \citep{Atek:2015},  which is almost two orders of magnitude dimmer than the faintest galaxies observed in random fields \citep{Bouwens:2014, Schenker:2013, McLure:2013}. Along similar lines,   \citet{Alavi:2014} used strong lensing by a foreground galaxy cluster to detect $z\sim 2$ galaxies which are two orders of magnitude fainter than what is normally observed at this redshift.  Because  strong lensing is so useful for the detection of faint high redshift sources, its statistics and properties were widely studied in literature (e.g., by \citet{Wyithe:2011} and references therein). However, strong lensing events are relatively rare with a raw probability for multiple images at the highest source redshifts of only $0.5\%$ \citep{Barkana:2000, Comerford:2002}.  On the other hand, galaxies which are not strongly lensed still undergo magnification (or de-magnification) by  foreground structure and can be mildly magnified  without being multiply imaged.   In this case the effect of lensing can be easily overlooked leading to an over (or under) prediction of the number counts of background galaxies, and thus, to an erroneous estimation of their properties. The effect of the intermediate and small magnifications on the observed properties of high redshift sources, such as their luminosity function, has not been properly addressed in the literature and is considered in this work for the first time.  However, the importance of lensing events with intermediate magnifications is being realized; for instance, \citep{Mason:2015} mention the relevance of newly detected at $z\sim 8$ sources with $1.4<\mu<2$ for the determination of the intrinsic luminosity function.

Here we explore the effect of gravitational lensing  by two types of foreground populations: (i) virialized halos hosting bright galaxies, and (ii) proto-clusters, i.e., non-virialized mildly non-linear overdensities, on the luminosity functions of the sub-mm and Lyman-break galaxies. In \S 2 we outline the basics of our lens model and compute the probability for lensing at each magnification, $P(\mu)$, which is necessary for deriving the observed luminosity functions. In \S 2.1 we carefully examine the case  when the background sources are  LBGs and  lensing is done by virialized halos which host bright galaxies paying particular attention to the surface brightness of the source and the lens galaxies. In this setup the source images could be hidden behind an extended lens that appears to be bright in the observable range of wavelengths, which, as we find, has a crucial effect on the lensing statistics. In \S 3 we present our results,  showing that if not accounted for,  gravitational lensing with intermediate and small magnifications ($\mu \la 2$) can be responsible for errors in the derived parameters of the observed source luminosity function of the two populations. Finally, we conclude in \S 4. Throughout this work, we adopt the standard set of cosmological parameters \citep{Ade:2014}.

\section{Lensing Model}

A central ingredient of our calculation is the probability for lensing with each given magnification, $P(\mu)$. We analytically compute this quantity  largely following the approach taken by  \citet{Lima:2010b}, whose main assumption is that for each trajectory of a photon emitted by a source there is a single massive object which plays a dominant role in the photon's deflection and  magnifies the background light by a factor of $\mu$ through gravitational lensing.   In the cases when multiple images are expected to form we keep only the stronger magnified image motivated by the fact that it is easier to observe. 

Our computational approach can be summarized as follows. First, for a lens hosted by a halo of mass $M_l$ and a source hosted by a halo of mass $M_s$ located at redshifts $z_l$ and $z_s$ respectively  the magnification at each impact parameter in the lens space, $r$,  is computed. Next, running over all possible parameters that describe the lens-source system, one finds the fraction of the parameter space, $f_\mu$, which yields magnifications higher then $\mu$  considering only the stronger magnified image in the case when multiple images are expected to form. Finally, the  lensing statistics is constructed by computing  the probability for lensing with magnification larger than $\mu$,  $P(>\mu) = 1-e^{-f_\mu}$, and then the probability for lensing with each given magnification, $P(\mu) = -dP(>\mu)/d\mu$, is calculated, while the total flux is conserved  ($<\mu> = 1$) and $P(\mu)$ is normalized to unity over the entire range of magnifications (see  \citet{Lima:2010b} for complementary details).

The factor $f_\mu$ can be simply written as,
\begin{align}
& f_\mu = \int_{0}^{z_s} dz_l\frac{D_A^2(z_l)}{H(z_l)}\int d\log M_l \frac{dn}{d\log M_l}\nonumber\\
& \int  \frac{d\log M_s}{N}\frac{dn}{d\log M_s} \Delta\Omega_\mu(M_l,z_l,M_s,z_s),
\label{Eq:fmu}
\end{align}
where $\Delta\Omega_\mu(M_l,z_l,M_s,z_s)$ is the angular cross-section for lensing  with magnification larger than $\mu$. If the redshift distribution of sources is given, we should average $f_\mu$ over $z_s$ as well, as we do in the case of sub-mm  galaxies taking their redshift distribution from the work by \citet{Simpson:2014}.  In equation (\ref{Eq:fmu}) $D_A$ is the angular diameter distance, and $H(z)$ is the Hubble constant at redshift $z$. The comoving number density of objects, $ dn/d\log M $ and the normalization factor,  $N = \int d\log M_s dn/d\log M_s$, are found using the Sheth-Tormen mass function \citep{Sheth:1999}. To correctly estimate the number density, $dn$, of the objects with masses between $M$ and $M+dM$ per a logarithmic mass interval, $d\log M$, one needs to specify the variance of fluctuations in the matter density at each mass scale (which we calculate using the outputs of CAMB\footnote{http://camb.info}), and the critical overdensity at which the objects form, $\delta_c$. The latter quantity depends on the type of the objects for which we want to know the number densities. As noted above, in this work we are interested in the effects produced by two kinds of foreground populations: (1)  virialized massive halos for which the critical overdensity is just the standard value for collapse (e.g., \citet{Barkana:2001}), and (2) non-virialized objects. To model the population of non-virialized objects we  rely on the fact that the radius of  a growing overdensity is  close to its value at turnaround (the  moment at which, in the framework of the spherical collapse model, the external mass shell has zero velocity) during most of the evolution time of the overdense region. We then find the critical overdensity, $\delta_c$, for the objects at turnaround by consistently solving the spherical collapse model with relevant cosmological parameters  at each redshift, and plug it into the Sheth-Tormen prescription to get the number density of non-virialized halos.

Next we calculate  the value of  magnification at each  impact parameter, which depends on the three-dimensional  density profile of each lens as well as on its distance from the source and the observer. In treating objects at turnaround we consider their mean density in the context of the  spherical top-hat collapse model. We  follow the standard procedure (e.g., outlined by \citet{Lima:2010b}) to find the magnification pattern for each overdensity. In general, the  objects at turnaround are only mildly nonlinear  having the  overdensities of the order $\sim 5$ (with the exact value being redshift dependent in $\Lambda$CDM)  which results in magnification of $\mu\la 2$.  Such objects do not produce multiple images and thus can have only a mild effect on the observable population.

In the case of virialized halos the model is more complicated. Conventionally, either the \citet{Navarro:1997} (NFW) or the  singular isothermal sphere (SIS) profiles are used to model the three-dimensional  density distribution. However, neither of the two profiles describes lensing by realistic galaxies close enough. In particular, observations of flat rotational curves in massive galaxies show that the mass distribution follows the SIS profile within 10-20 inner kpc \citep{Kochanek:1994}, while the outskirts are better described by the NFW profile \citep{Mandelbaum:2005}.  Therefore we adopt a combination of SIS and NFW through a piecewise form for the magnification as a function of the impact parameter in the lens plane
\begin{equation}
\mu =\left\{
\begin{array}{l}
\mu_{SIS}, ~~~r<r_c\\
\mu_{NFW},~~~r>r_c
\end{array}\right.
\end{equation}
and connect the two profiles at a projected radius $r_c$ that characterizes the extent of the luminous core of the galaxy (twice the half light radius). 

The last ingredient needed for equation (\ref{Eq:fmu}) is the angular cross-section for lensing, $\Delta\Omega_\mu(M_l,z_l,M_s,z_s)$. To calculate this quantity for each lens-source pair  we go over all $r<r_0$, where $r_0$ is the maximal value of the impact parameter that allows a single lens to dominate.  In the case when multiple images are produced we consider the $r$ which corresponds to the brightest image only. By averaging over all the masses of lenses at each redshift, we first calculate the mean separation between halos, $\bar s$, in terms of either the mean virial radius (averaged over halos masses at each redshift) or the typical radius of a structure that is turning around. In the former case we find $\bar s \sim 10$, while in the latter case  $\bar s \sim 2.5$, with the exact value in each case being redshift dependent.  Next, we assume that each given object dominates the lensing effect for  the impact parameters smaller than $\bar s $  times its radius, for which we adopt either a virial radius, and thus $r_0 \sim  10 ~r_{vir}$, or a radius at turnaround, $r_0 \sim 2.5~ r_{ta}$, depending on the case studied. 
  
\begin{figure*}
\includegraphics[width=3.4in]{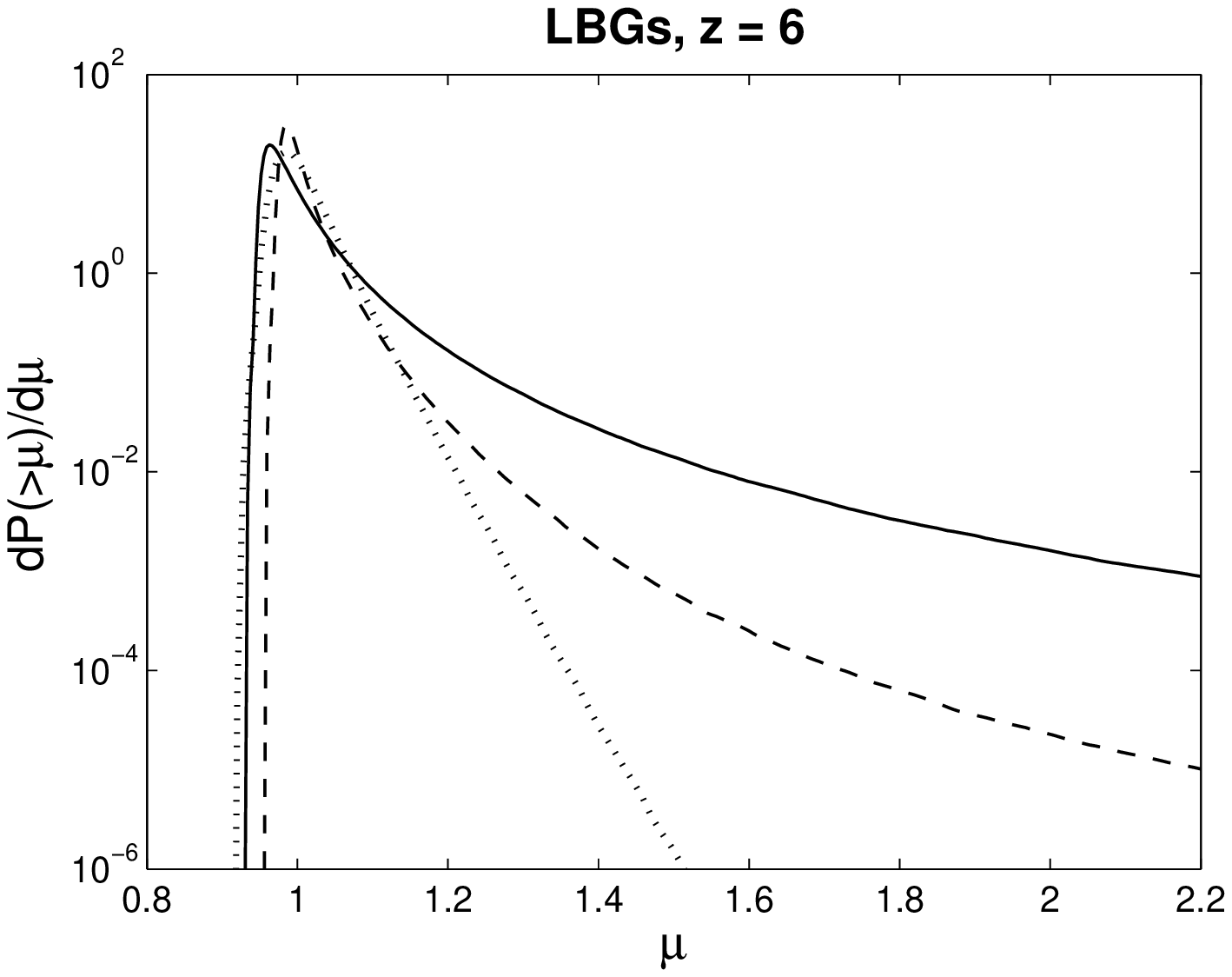}\includegraphics[width=3.4in]{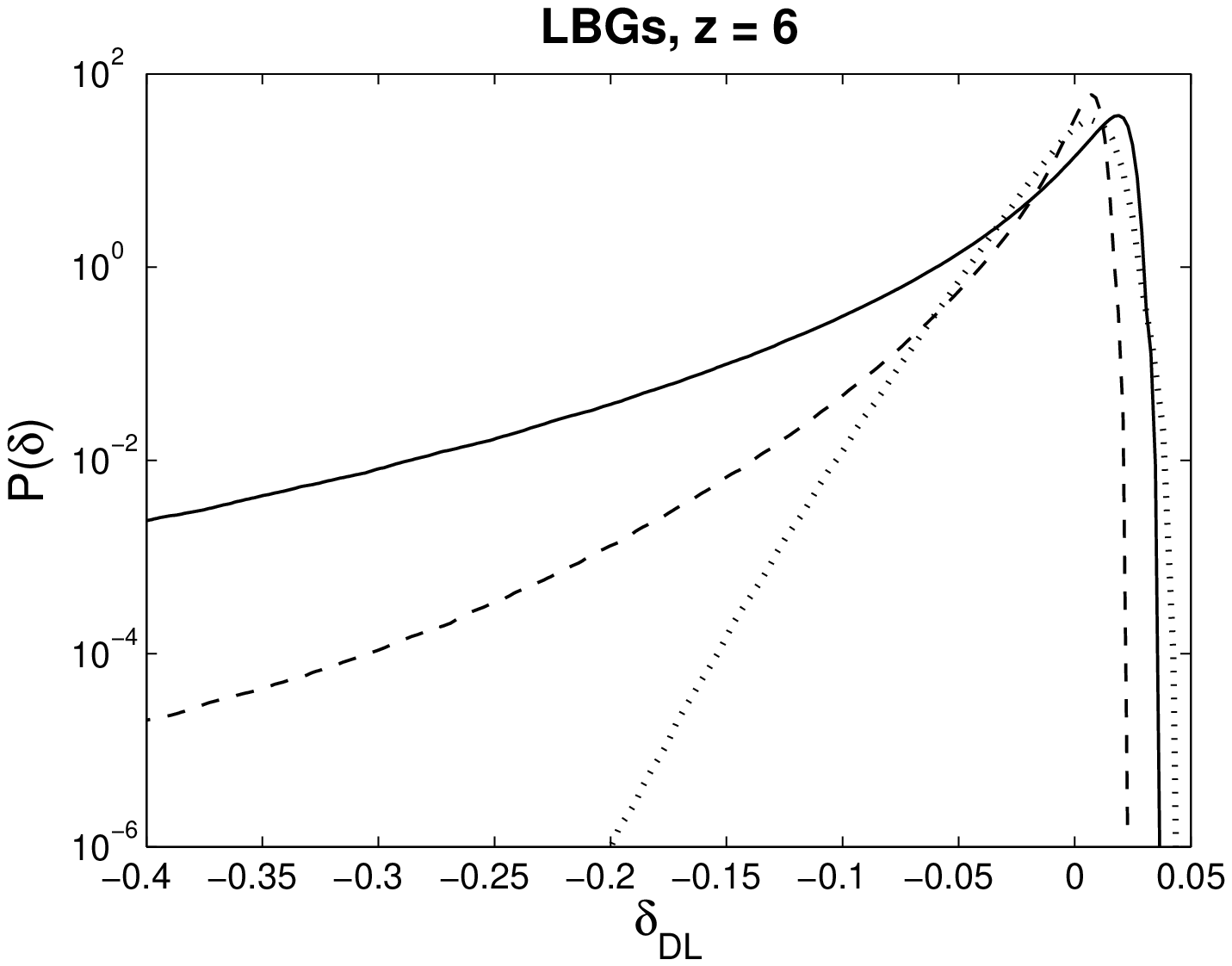}
\includegraphics[width=3.4in]{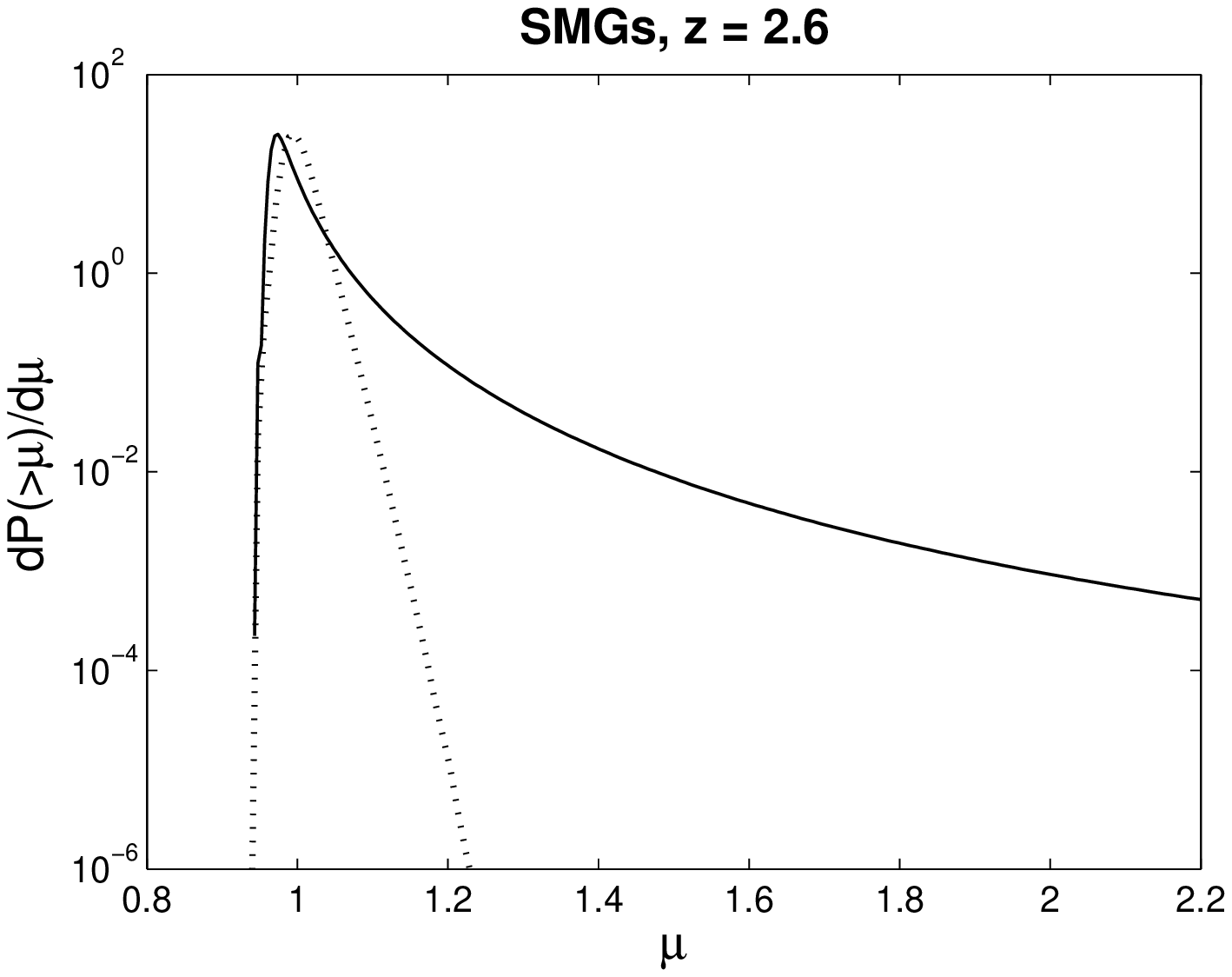}\includegraphics[width=3.4in]{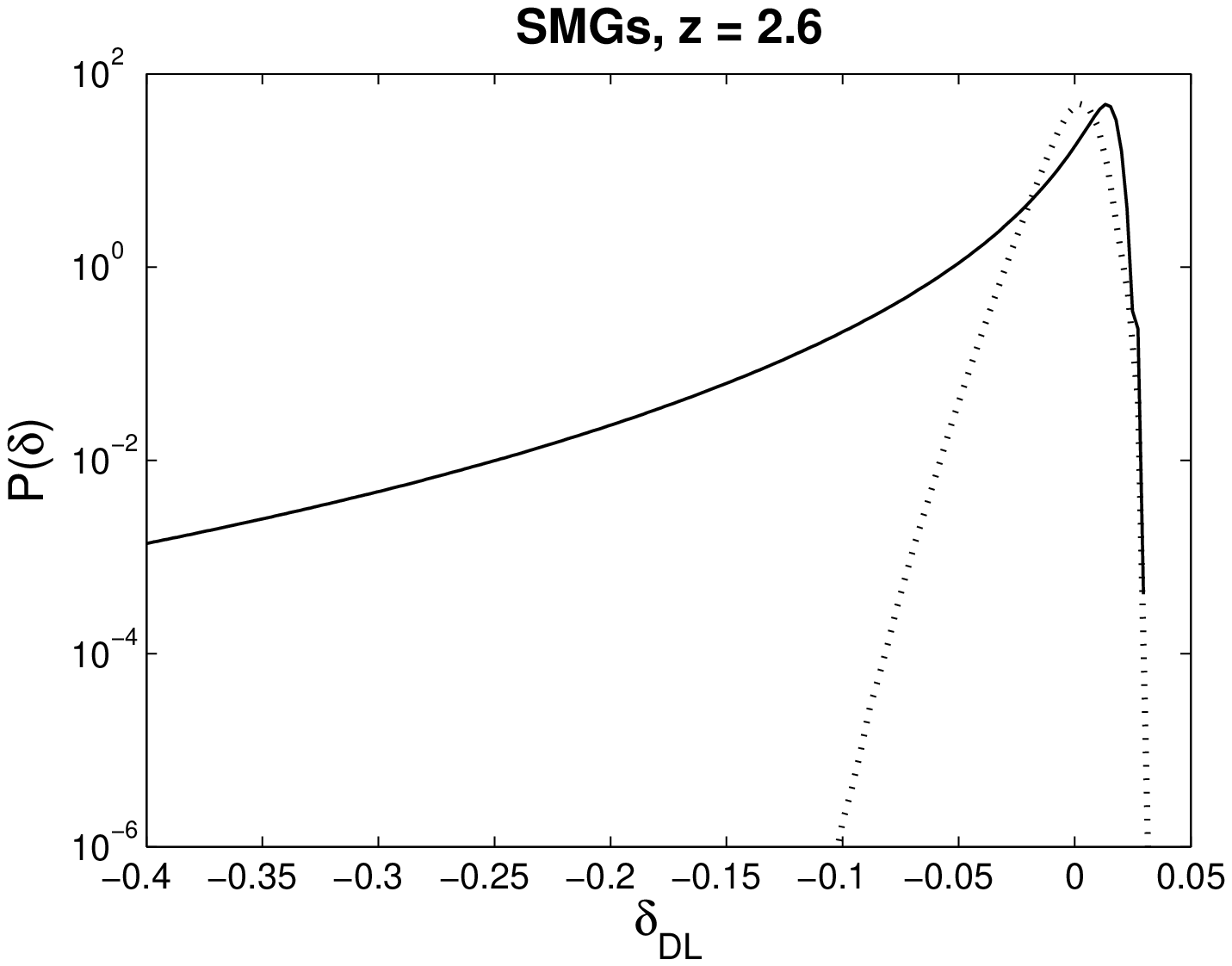}
\caption{Probability distribution for lensing with magnification $\mu$ (left panel) and probability distribution for $\delta_{DL}$ (right   panel) shown for a population of the Lyman-break galaxies (LBGs) at $z_s = 6$ (top)  and of sub-mm galaxies (SMGs) at $\bar z_s = 2.6$ (bottom). For  SMGs, the lenses under consideration are  virialized halos (solid) and proto-clusters (dotted). In the case of LBGs the sources are lensed by a population of virialized galaxies including the reduction in $P(\mu)$ (dashed) and   ignoring it (solid), as well as by proto-clusters   (dotted). }
\label{fig:pdfs}
\end{figure*}

The resulting $P(\mu)$ is shown on the left panels of Figure \ref{fig:pdfs}  for the two source populations, i.e.,  LBGs at $z_s = 6$ and for SMGs at $\bar z_s = 2.6$ (which represents the center of the current redshift sample of SMGs), and for the two lens populations, i.e., for virialized and non-virialized halos. For LBGs, a particular care should be taken when computing the lensing statistics since the foreground galaxies are normally bright in the band which refers to the rest-frame UV of the LBGs. This effect introduces a suppression in the probability for lensing at strong and intermediate magnifications, which we also show on the Figure, and discus in full detail in \S 2.1.

For both sets of sources, LBGs and SMGs, and in the case when lensing is due to the  highly nonlinear overdense regions (i.e., virialized halos), the large magnification tail of $P(\mu)$ scales as $\propto \mu^{-3}$ as expected \citep{Turner:1984}, and all values of magnifications can be obtained,  including very large $\mu$ for which Einstein rings are produced. On the contrary, in the case when lensing is due to the  mildly non-linear objects,  $P(\mu)$ drops very fast with magnification, not allowing for the possibility of strong lensing at all.  However, as we see in \S 3, the effect of the objects at turnaround on the bright end of the luminosity function can be comparable to lensing by halos when observing a sample of field galaxies which do not experience strong lensing. 

 The effect of gravitational lensing can also be interpreted as modification of the luminosity distances. On the right panels of Figure \ref{fig:pdfs} we show the probability distribution for the fractional change in the luminosity distance due to the magnification of the flux, $\delta_{DL} = \mu^{-1/2}-1$. Although the probability distribution of $\delta_{DL}$ peaks around zero (i.e., magnification has no effect in average around the sky) there is dispersion in the values of the luminosity distances simply because along a random line of sight the flux can be magnified by a random factor $\mu$. This dispersion in the luminosity distance of high redshift objects should be carefully accounted for. In particular, the precision with which cosmological parameters from a high-redshift sample of standard candles can be determined is expected to be affected by gravitational lensing. However, a more quantitative determination of the precision in the cosmological parameters measured using the high-redshift standard candles, as well as the effect of lensing,   is beyond the scope of current paper.

\subsection{Reduced lensing probability for the LBGs behind a bright lens}

Even when a source is located behind a bright lens, it can still be  separated based on its different colors \citep{Barkana:2000}. Here we apply another argument, showing that if  magnified enough, the  source can outshine the foreground galaxy and be observable even when located behind its extended luminous core, i.e.,  surface brightness may be larger than that of the lens for some values of the impact parameter.  This argument introduces a new criterion, previously ignored in the discussions of the galaxies' luminosity function, which is relevant in the case of the high-redshift LBGs. Since foreground galaxies are normally bright in the observed bands which correspond to the rest frame UV bands of the sources, some images are too faint to be seen through the bright part of the lens even when they are magnified. As a result, the probability to observe LBGs with intermediate and large magnifications is reduced, as we show below.  

In our analysis we relate to each massive halo an exponential surface brightness profile with a half light radius $r_c/2$ \citep{Szomoru:2013,   Kravtsov:2013} and a total UV luminosity $L_{UV}$. The UV luminosity of each halo is directly proportional to the star formation rate in the halo, ($L_{UV}$/erg s$^{-1}$ Hz$^{-1})$ $\approx$ $7\times10^{27}$ (SFR/M$_\odot$ yr$^{-1}$) \citep{Madau:1998}, which in turn scales with its mass and redshift \citep{Behroozi:2013, Behroozi:2014}.

The lensing kernel for LBGs at $z_s\sim 6-10$ is dominated by contribution from lenses at $z_l\sim 1$, thus  implying that the emission wavelength of $z_s\sim 10$  sources is almost an order of magnitude shorter than that of the lens galaxy. When the high-redshift sources are observed at the rest-frame wavelength of 1500 A (e.g, by WFC3 camera on board HST), the foreground galaxy would frequently be observed at the wavelengths bluer than the Balmer jump. Ignoring lines, the spectral energy distribution (SED) of galaxies at the corresponding wavelength band is expected to be rather flat. We use a toy model to include the SED of foreground galaxies and account for K-corrections.  Our toy model for the galaxy SED consists of a power-law with the observed slope according to \citet{Kurczynski:2014} for wavelengthes within UV-continuum.  We model the Balmer jump adopting observations by \citet{Oteo:2014}, and assume a flat spectrum at longer wavelengthes. A more detailed treatment of  K-corrections in a more solid way, e.g. using SED models provided by  \citet{Bruzual:2003}, goes beyond the scope of this paper.

\begin{figure*}
\includegraphics[width=3.4in]{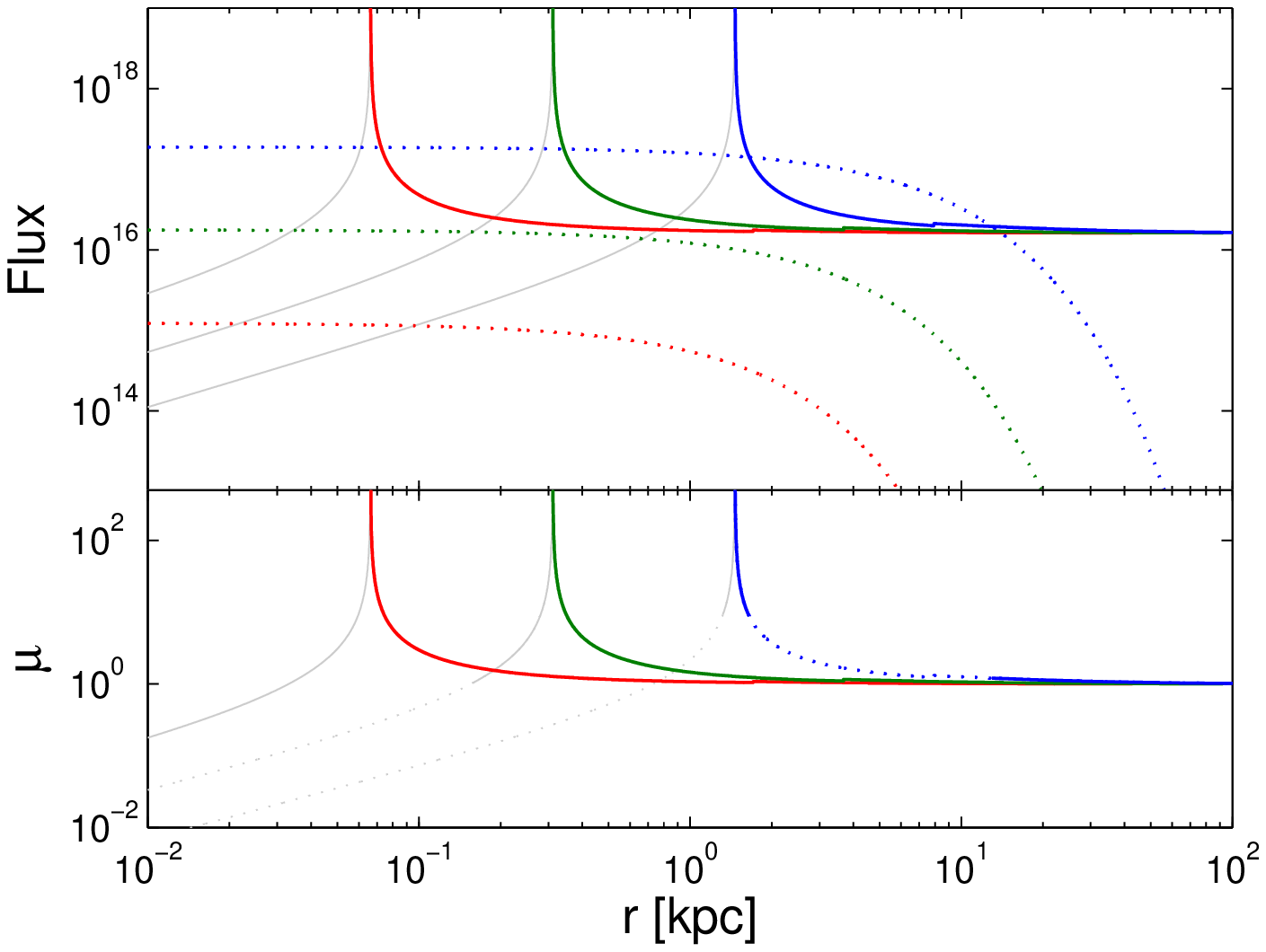}\includegraphics[width=3.4in]{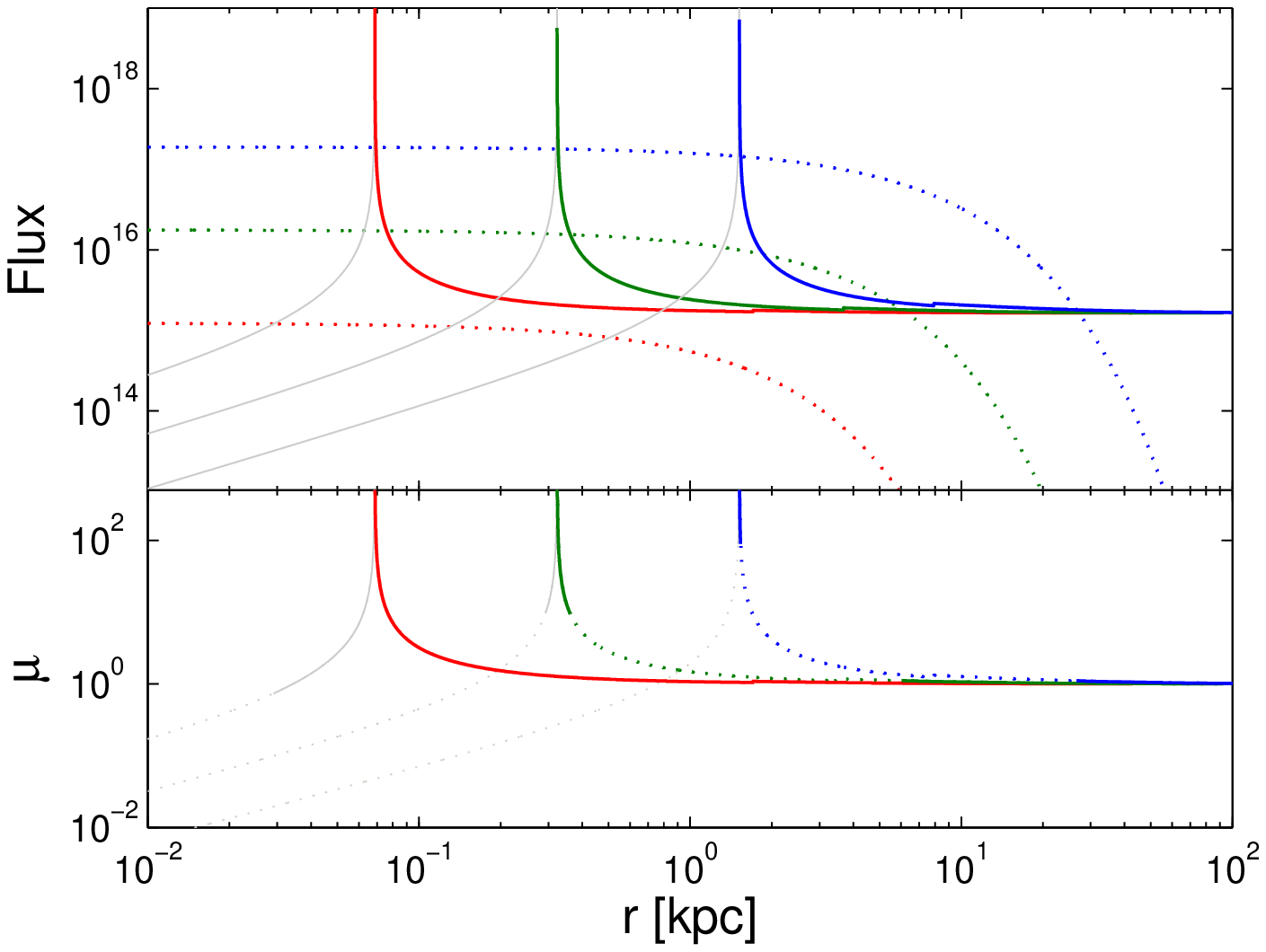}
\caption{The top panels show the fluxes (in erg s$^{-1}$ Hz$^{-1}$ kpc$^{-2}$ units) of foreground galaxies (dotted   lines) and the fluxes of a background galaxy magnified by each lens  (solid lines) as a function of the physical impact parameter
  in the lens plane. Left: Source of mass $M_s = 10^{11}$
  M$_\odot$ at redshift $z_s = 6$.  Right: Source of mass $M_s
  = 5\times 10^{9}$ M$_\odot$ at redshift $z_s = 10$. In each case
  three types of foreground galaxies at $z_l = 0.5$ are shown: $M_l =
  10^{10}$ M$_\odot$ (red), $M_l = 10^{11}$ M$_\odot$ (green), and
  $M_l = 10^{12}$ M$_\odot$ (blue).  The bottom panels show the
  corresponding magnification in each case in the observable part
  (solid lines) and non-observable part (dotted lines).   Light grey lines  correspond to the weakly magnified image in each case and are shown for completeness.}
\label{fig:lum}
\end{figure*}

Using all the model ingredients described above we calculate the observed flux per unit frequency for both the source and the lens including  the magnification of the source. This allows us to find impact parameters (and magnifications) for which the source surface brightness is higher than that of the lens. Such regions of the parameter space contribute to the lensing statistics. In Figure \ref{fig:lum} we show the observed fluxes for several lens-source pairs accounting also for the magnification of the background galaxy by the foreground galaxy. Specifically, we show the cases of two sources $M_s = 10^{11}$  M$_\odot$ at $z_s = 6$ (left panel) and  $M_s = 5\times 10^{9}$  M$_\odot$ at $z_s = 10$ (right panel) lensed by three different lenses (M$_l = 10^{10}$ M$_\odot$,  M$_l = 10^{11}$ M$_\odot$  and  M$_l = 10^{12}$ M$_\odot$) at $z_l=0.5$. The masses of the deflectors were chosen to cover the most typical (M$_l = 10^{10}$ M$_\odot$) and massive (M$_l = 10^{12}$ M$_\odot$) structures at that epoch. Our choice of $M_s$  is motivated by the derived masses of the halos hosting Lyman-break galaxies recently reported by \citet{Barone:2014}. In some cases, e.g., for an M$_l = 10^{12}$ M$_\odot$ lens at $z_l = 0.5$ and a source of M$_s = 10^{11}$ M$_\odot$ at $z_s = 6$ (blue curve on the left panel of Figure \ref{fig:lum}), there is a range of impact parameters for which the magnified source is brighter than the lens and thus can be observed.  In this particular case, the range of impact parameters at which  the image  is visible includes  $r\geq 13$ kpc and $1.3 \leq r \leq 1.6$. Note that the range of $r$ smaller than the radius of the Einstein ring (here $r_E = 1.46$ kpc) shown with light grey curves on the Figure, is attributed to the weakly magnified images and is not included in the calculation of $P(\mu)$ plotted in Figure \ref{fig:pdfs}. This means that for $r\geq r_E$ the range of magnifications between $\mu = 8.8$ at $r=1.6$ kpc and $\mu = 1.2$ at $r = 13$ kpc  is not observable and such magnifications do not contribute to our lensing statistics. The right panel of Figure \ref{fig:lum}  shows that  this phenomenon is essentially the same for a different choice of the mass and redshift of the source. However, the precise range of the impact parameters within which the magnified source can be observed in each case depends on the masses and redshifts of both the lens and the source. 

The fact that in some cases (for some values of the impact parameter) the foreground galaxies outshine the background LBG sources affects statistics of gravitational lensing, as shown in Figure \ref{fig:pdfs}.   Within the outlined setup, probability of obtaining any magnifications ($1<\mu < \infty $) is affected by this selection criterion, as can be seen from the Figure where  $P(\mu)$ is shown with and without such a reduction; however, the slope of the probability distribution at large magnifications $\mu > 2$ (strong lensing regime) is not strongly affected by this effect and scales  as $\propto \mu^{-3}$. As shown in the next section, the reduction in $P(\mu)$ affects the observed luminosity function of LBGs at its bright end.

\section{Luminosity Function of High-redshift Field Galaxies}

To account for the effect of gravitational lensing on the luminosity function, one must convolve the differential probability $P(\mu)$ which we found in \S2 with the intrinsic luminosity function of the population under consideration, $\Psi_{int}(L)$. The observed luminosity function,
\begin{equation}
\Psi_{obs}(L_{obs}) = \int^{\mu_{max}}_{\mu_{min}} \frac{P(\mu)}{\mu}\Psi_{int}(L=L_{obs}/\mu)d\mu,
\label{Eq:LFobs}
\end{equation}
is modified by lensing and depending on $P(\mu)$ may have a completely different shape than the intrinsic one.

The effect of lensing on the observed properties of sources depends on the shape of the intrinsic luminosity function of the population under consideration. In particular, if $\Psi_{int}(L)$ is flatter than $L^{-3}$, then lensing has only a minor effect on the observed luminosity function \citep{Blandford:1992}. Here we apply our findings to the LBGs  at redhifts $z_s\gtrsim6$ and to the SMGs at $\bar z_s = 2.6$.  Although rather steep, the faint-end luminosity function of each one of these two populations appears to be  flatter than  the critical dependence, i.e., $\Psi(L)\propto L^{-3}$. Therefore lensing is not expected to have any  impact on the number counts of intrinsically faint sources. However, at luminosities higher than a characteristic value, $L^*$ (or at the fluxes larger than a characteristic value $S^*$), the luminosity function is expected to drop rapidly following either an exponential or a steep power-law.  In this regime the effect of lensing is dramatic: it boosts the observed number counts of the brighter sources by magnifying the luminosities of intrinsically fainter ones. Since the abundance of intrinsically faint galaxies is much higher than that of the bright ones, the increase of the observed number counts at the bright end is striking. Specifically, the exponential drop of the Schechter function of LBGs transforms into a power-law with the index of -3 when lensed \citep{Wyithe:2011}.

At present, the bright end of the luminosity function is not very well constrained for both populations that we consider here due to the substantial cosmic variance \citep{Newman:2002, Somerville:2004, Trenti:2008}. In the particular case of the LBGs at $z_s\sim 9-10$, the data from GOODS-N or GOODS-S WFC3/IR surveys (with the total area of $\sim  150$ arcmin$^2$) suffer from the cosmic variance of $ 15 - 20\%$ on the overall number counts out of which the luminous sources beyond $L^*$ make only a small fraction \citep{Oesch:2014}. Similarly, in the case of $z_s\sim 6$ LBGs, counted in 5 independent $20′′\times 7.5′′$ CANDELS survey fields, a total uncertainty of 10$\%$ on the volume density of galaxies from cosmic variance is estimated \citep{Bouwens:2014}. In the case of the SMGs, the luminosity function is poorly understood at all the luminosities,  and its  redshift evolution and $S^*$ remain unconstrained\footnote{ Although several studies provided the luminosity function of SMGs based on the observations  by the  {\it Herschel} Space Observatory \citep{Lapi:2011, Gonzalez-Nuevo:2012}, the quoted number counts at the bright-end may be overestimated as a result of the poor angular resolution of the satellite. Recent studies with ALMA, which has much higher angular resolution and is able to resolve the SMGs at high redshifts, show that some of the previously unresolved bright sources are, in fact, groups of highly clustered dimmer galaxies \citep{Bussmann:2015}.}.  The first detailed surveys in the sub-mm range are currently on their way, for example with  ALMA, and are expected to provide better constraints on the number counts of SMGs.  The effect of lensing on the luminosity functions of both LBGs and SMGs may appear to be significant  once  these populations  are better explored. In the following we analyze the effect of lensing on the luminosity functions for all possible magnifications, while paying particular attention to the impact of magnifications in  the intermediate range $\mu\la2$, which is expected to affect most of the galaxies in the field, and comparing it with the case of strong lensing. 

\subsection{LBGs}

The luminosity function of high-redshift galaxies is commonly fitted by the Schechter form \citep{Schechter:1976, Oesch:2014,  Bouwens:2014}
\begin{equation}
\Psi(L)= \frac{\Psi^*}{L^*}\left(\frac{L}{L^*}\right)^\alpha\exp\left(-\frac{L}{L^*}\right).
\label{Eq:LFint}
\end{equation}
It is characterized by a power law dependence with index $\alpha$ at the faint-end and an exponential drop at luminosities higher than the critical value $L^*$, where $\Psi^*$ determines the overall normalization of the number counts. As discussed above, the abundance of rare sources at the luminous end of the luminosity function, $L\gg L^*$, is very sensitive to lensing due to the steepness of $\Psi(L)$, which cuts-off exponentially. 

The effect of strong lensing on the luminosity function of the Schechter form  for high-redshift sources has been extensively studied in  literature (see e.g., \citet{Wyithe:2011}). It was shown that in the presence of significant gravitational lensing, the luminosity function acquires a power-law slope with an index $-3$ at its bright-end, so that the apparent abundance of galaxies with $L>L^*$ is dramatically enhanced. Observing such a strongly distorted luminosity function would flag strong lensing. However, only a small part of high-redshift galaxies are strongly lensed, while the vast majority of galaxies experiences either intermediate or weak lensing.  

\begin{figure*}
\includegraphics[width=3.4in]{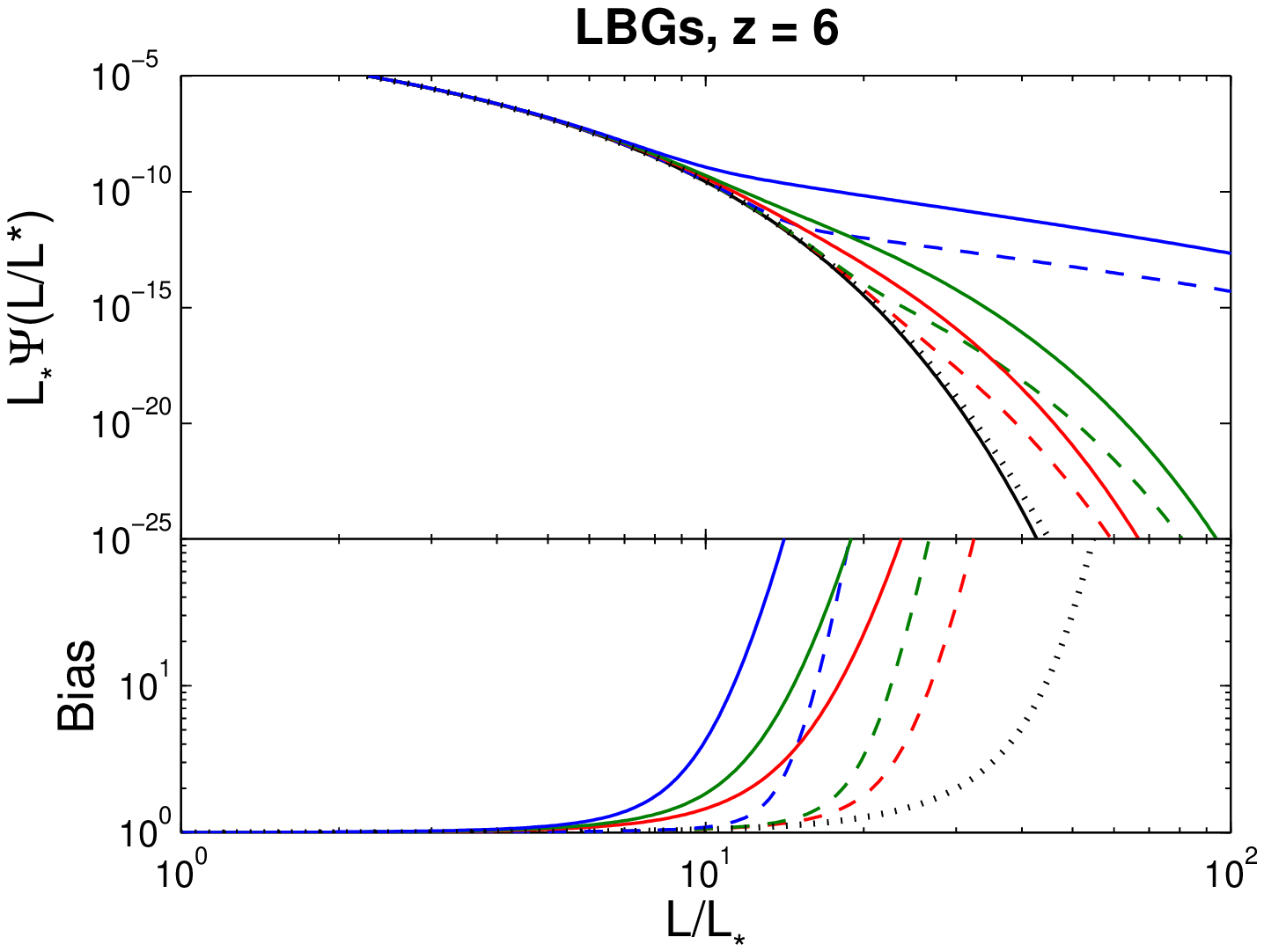}\includegraphics[width=3.4in]{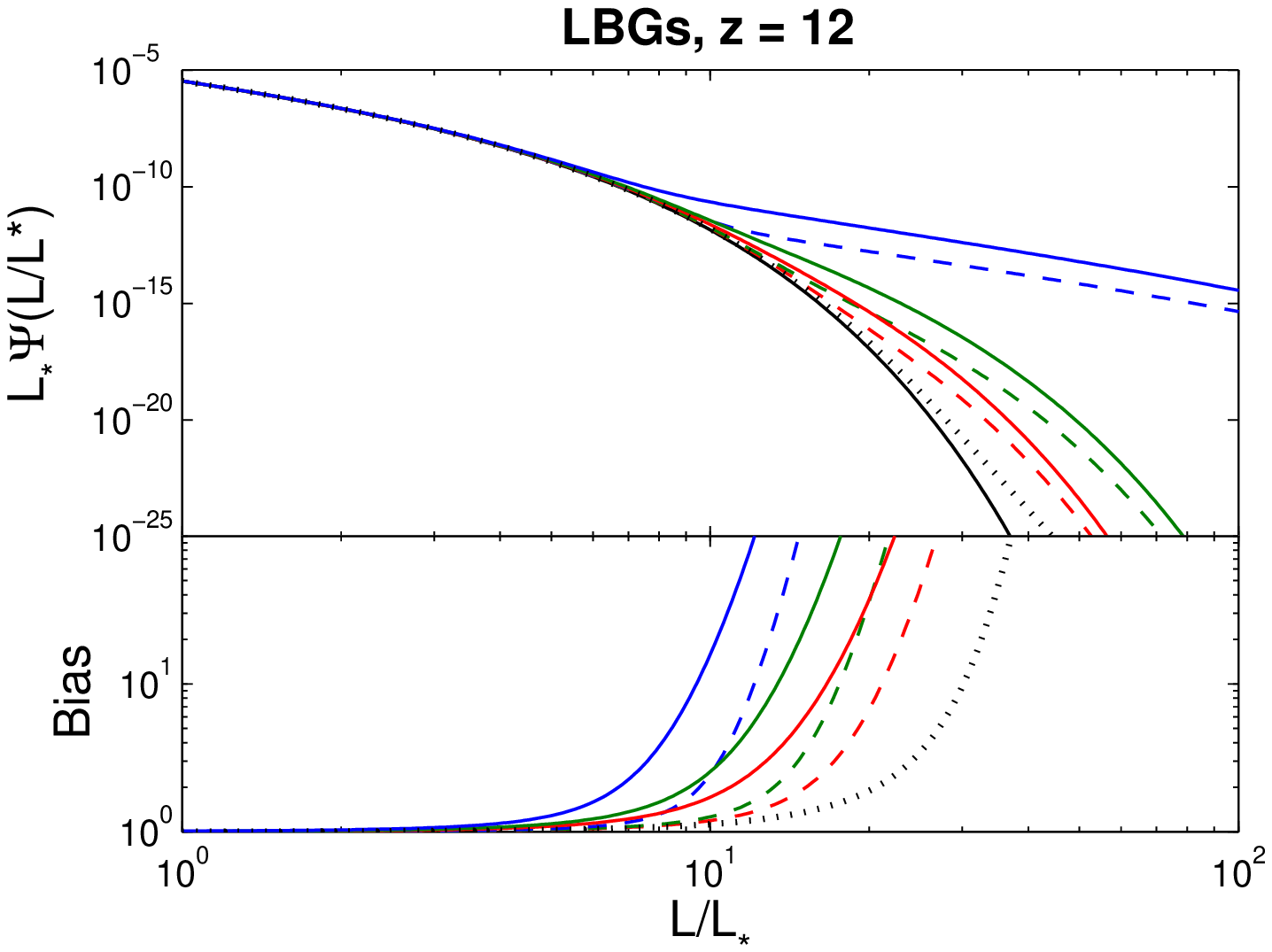}
\caption{ The luminosity function multiplied by $L^*$  (top panels, $L^*\Psi$ is shown in units of Mpc$^{-3}$ and $L^*$ is calculated in erg s$^{-1}$ Hz$^{-1}$) and the magnification bias,  $\Psi_{obs}(L)/\Psi_{int}(L)$, (bottom panels)  of the LBGs  at $z_s = 6$ (left) and $z_s = 12$ (right). We show the results for the intrinsic luminosity function (black line), and the luminosity function of LBGs  lensed by  a population of virialized objects including the reduction in $P(\mu)$ for all possible magnifications (dashed) and   ignoring it (solid) for magnifications $\mu\leq 2$ (red),  $\mu\leq 3$ (green),  $\mu\leq 1000$ (blue) as well as for LBGs lensed by non-virialized objects at turn around   (black dotted). }
\label{fig:LF}
\end{figure*}

We start by quantifying the effect of the ``common''  lensing on the luminosity function of LBGs. Adopting a  set of Schechter parameters found by \citet{Bouwens:2014} for LBGs observed at $z_s = 4-10$ ($\Psi^* =   0.44\times10^{-0.28(z_s-6)}10^{-3}$ Mpc$^{-3}$, $\alpha = -1.87-0.1(z_s-6)$, and $M_{UV}^* =  -20.97$ where the UV magnitude is related to luminosity through $M_{UV}^*\propto  -2.5\log_{10 } L^*$), we apply our calculated $P(\mu)$ to estimate the observed luminosity function and bias, defined as   a ratio of lensed to unlensed luminosity functions at a given luminosity, $\Psi_{obs}(L)/\Psi_{int}(L)$, which are shown on  Figure \ref{fig:LF} for several cases of maximal magnification $\mu_{max}$ at two redshifts (6 and 12). In particular, we show  $\mu_{max} = $ 2, 3, and 1000, where the last case includes the effect of strong lensing with Einstein rings, including and excluding the reduction in $P(\mu)$ of LBGs at intermediate magnifications  discuses in \S 2.1. Finally, we show the case of lensing by non-virialized halos at turn-around.

\begin{deluxetable*}{rcccccccc}
\tabletypesize{\scriptsize} \tablecaption{ Summary of the errors introduced by gravitational lensing in the luminosity function of LBGs at different redshifts. Each   entry contains the pair of values:  (Bias measured at $L = 26 L^*$, $\Delta \Psi^*\%$) introduced by lensing with $\mu_{max} =$ 1.3, 1.5, 1.7, 2 and 3 due to virialized halos in the parameters of the Schechter luminosity function. We quote the results including the reduction in $P(\mu)$ (the upper set of numbers in each case) and excluding it (the lower set).  In addition, we show the case of  the lensing by proto-clusters with all possible magnification (the last column). Fits are done for the  luminosities in the range of magnitudes larger than the detection limit of HST   (for sources at $z_s\leq10$) and JWST (for $z_s>10$) and for the bright magnitude   limit $M_{UV} = -24.5$ corresponding to $L_{max} \approx 26 L^*$.} \tablewidth{0pt}
\tablehead{ \colhead{Redshifts} & \colhead{$\mu_{max} =    1.3$}&  \colhead{$\mu_{max} =
    1.5$}& \colhead{$\mu_{max} =    1.7$}&\colhead{ $\mu_{max} =2$} & \colhead{$\mu_{max} =3$} & \colhead{$\mu_{all}^{ta} $}}
\startdata 
6 & 1.7, 1.6$\%$ &  2.4, 2.3 $\%$ & 3.6, 2.8$\%$& 7.2, 3.8$\%$& 62.1, 7.4$\%$& 1.6, 1.6$\%$\\ 
 & 4.2, 3.8$\%$ &  16.6, 8.6 $\%$ & 58.2, 13.0$\%$& 266.2, 18.9$\%$& $>10^3$, 33.7$\%$& \\ 
\hline 
8 &  2.0, 2.1$\%$& 3.5, 3.3$\%$&  6.4, 4.2$\%$& 15.7, 5.9$\%$ & 181.2, 11.2$\%$ & 2.1, 2.7$\%$\\ 
 &  4.6, 3.9$\%$& 19.0, 9.1$\%$&  69.3, 14.0$\%$& 329.7, 20.5$\%$ & $>10^3$, 36.5$\%$ & \\ 
\hline 
10 &  2.4, 2.6$\%$& 5.0, 4.5$\%$& 10.8, 5.9$\%$& 32.8, 8.4$\%$ & 467.6, 15.6$\%$ & 2.8, 3.3$\%$\\
 &  4.8, 4.0$\%$& 21.5, 9.9$\%$& 77.9, 15.1$\%$& 401.3, 22.2$\%$ &$>10^3$, 39.2$\%$ &\\ 
\hline 
12 & 2.8, 3.0$\%$& 7.1, 5.7$\%$& 17.9, 7.9$\%$& 64.5, 11.4$\%$ & $>10^3$, 20.8$\%$ & 4.0, 4.5$\%$\\ 
 & 5.2, 4.2$\%$& 23.3, 10.4$\%$& 90.4, 16.1$\%$& 460.4, 23.3$\%$ & $>10^3$, 41.9$\%$ & \\ 
\hline 
15 & 3.7, 4.0$\%$& 12.1, 7.9$\%$ & 37.1, 11.4$\%$& 166.0, 16.4$\%$ & $>10^3$, 30.0$\%$ & 7.0, 6.0$\%$ \\ 
 & 5.4, 4.5$\%$& 27.0, 11.4$\%$ & 105.2, 17.5$\%$& 569.6, 25.6$\%$ & $>10^3$, 46.2$\%$ & \\ 
\enddata
\label{Tab:1}
\end{deluxetable*}

It is clear that the reduction in $P(\mu)$ plays an important role in determining the observed luminosity function of LBGs for all values of $\mu_{max}$. When measured at $L = 30 L^*$ in the cases of $\mu_{max} =$ 2, 3 and 1000, the ratios of the luminosity function which excludes the reduction in $P(\mu)$ to the one which includes it,  are 49, 109, and 59 at $z_s = 6$, while at $z_s = 12$ the numbers are 7.6, 12.5 and 9.7, respectively. In addition, it is important to note that mildly lensed population, e.g.,  with $\mu_{max}\leq 2$, can still be described by the Schechter function with slightly different $\Psi^*$ or $M_{UV}^*$,  if the latter parameter is allowed to float\footnote{\citet{Bouwens:2014} concluded that evolution of the   critical brightness $M_{UV}^*$ with redshift is insignificant and  its value is nearly constant (in terms of the UV absolute magnitudes   $M_{UV}^* = -20.97\pm 0.06$). However, the evolution of the   faint-end slope and normalization with redshift was found to be   significant, e.g., the steepening in the effective shape of the UV luminosity function was found to be significant at the $5.7 \sigma$ level   \citep{Bouwens:2014}.} (where we choose to keep $\alpha$ fixed since the faint end number counts should not be affected by lensing).  However, in fields which include stronger lensing events,  e.g., for $\mu_{max}\geq 3$, the shape of the luminosity function is significantly distorted. Even for $\mu_{max} = 3$ the Schechter form develops a ``secondary knee'', which (when stronger magnifications are included) transforms into the power law with the asymptotic index of $-3$ \citep{Wyithe:2011}. Visually the effect is similar for all the redshifts at which the luminosity function has been observed, $z_s\sim 4-10$ \citep{Bouwens:2014}. Lastly, the effect of lensing by non-virialized objects is small, but non-negligible. In fact, in the parts of the sky where there are no strong lenses it can be comparable to the lensing by virialized halos with $\mu_{max}\sim 1.4$ at all redshifts  (as we show below in Table \ref{Tab:1}).

Next, we quantify how strongly  lensing with $\mu_{max}\leq 3$ affects the fitting parameters of the Schechter function. When estimating the effect  on the luminosity function it is important to set the upper limit on the luminosity, $L_{max}$, which we expect to observe. This number depends on both the intrinsic distribution of high-redshift galaxies as well as on the volume of the survey. If the survey volume is too small, it will not  sample the brightest rarest sources. In particular, \citet{Oesch:2013} expect to find sources with apparent magnitudes in the range $26-30.5$ at $z_s\sim 10$ (which corresponds to the range from $M_{UV} = -17.0$ to $M_{UV} = -21.5$) using the Hubble Ultra-Deep Field. In this case $L_{max} = 1.6L^*$ and the data are not expected to be sensitive to lensing. Fitting the Schechter function to the lensed luminosity function with $\mu_{max} = 1000$ yields  discrepancy below $1\%$ in the value of $\Psi^*$ when compared to the  intrinsic case and  keeping $\alpha$ and $L^*$ fixed.  The effect of lensing on the parameters of the luminosity function starts to manifest itself when the brightest observed sources have $M_{UV} \leq -22.5$. In this regime, by comparing the strong lensing case to the no-lensing case we get a  $\sim 2\%$ discrepancy for LBGs at $z_s = 15$ in the value of $\Psi^*$ (with even a weaker effect on LBGs at lower redshifts), whereas the effect on the critical luminosity $L^*$ is still negligible. 
 When the brightest sources reach $M_{UV}\sim -23.5$, in which case $L_{max} \sim 10 L^*$, strong lensing starts to affect the shape of the luminosity function and the errors in fitting parameters are  $\Delta \Psi^*(\mu_{max} = 1000) \sim 6.0\%$, $\Delta \Psi^*(\mu_{max} = 5) \sim 2.3\%$ and $\Delta \Psi^*(\mu_{max} = 3) \sim 2.0\%$ for the source population at $z_s=15$ when the reduction in $P(\mu)$ is taken into account (and $ 30\%$,  $6.2\%$ and $4.2\%$ respectively when it is ignored); when the brightest objects are of  $M_{UV}\sim -24$ ($L_{max} \sim 16 L^*$) the errors  reach  $\Delta \Psi^*(\mu_{max} = 1000) \sim 30\%$, $\Delta \Psi^*(\mu_{max} = 5) \sim 11\%$ and $\Delta \Psi^*(\mu_{max} = 3) \sim 7\%$ with the reduction in $P(\mu)$ (and  $ 44\%$,  $21.6\%$ and $14.3\%$ when it is ignored). In the following we adopt the value $M_{UV} = -24.5$ for the brightest observable galaxy ($L_{max}/L^* \sim 26$) to put the upper limit constrains on the distortion of the luminosity function of LBGs due to lensing, and summarize the results in Table \ref{Tab:1}.  For this limiting  value of $M_{UV}$  we estimate the discrepancy in the normalization of the Schechter luminosity function (the discrepancy in $M_{UV}^*$ is  negligible reaching $\sim 2\%$ for $z_s = 15$ with $\mu_{max} = 3$) at a range of redshifts relevant for the observations with HST and JWST for  $\mu_{max} = $ 1.3, 1.5,  1.7,  2 and 3 including and excluding the reduction in $P(\mu)$. For higher values of $\mu_{max}$, i.e., when observing a field which includes strongly lensed (for example, the case of $\mu_{max} = 1000$ in Figure \ref{fig:LF}) bright sources ($L_{max}\sim 26 L^{*}$ and higher),  the effect of  lensing on the luminosity function is apparent and the resulting luminosity function cannot be fitted by the Schechter form.  Our predictions for the JWST redshift range ($z_s =  12$ and $15$) are based on the extrapolation of the results for the intrinsic luminosity function.  In addition, we quote the results for the lensing with proto-clusters for which the discrepancy in the Schechter parameters  appears to be comparable to the case of lensing by virialized halos with $1.3<\mu_{max}< 1.5$ at $z_s = 6-15$.  In agreement with Figure \ref{fig:LF}, the results in Table \ref{Tab:1} show that the effect of lensing is manifested stronger in the case when there is no reduction in $P(\mu)$ due to the surface brightness argument.  For instance, for sources at $z = 8$ the error in the normalization of the luminosity function drops from $\sim 21\%$ in the case when the surface brightness arguments are ignored  to $\sim 6\%$ when the reduction is accounted for. Therefore, the non-realistic case with no reduction in $P(\mu)$ can be used to set an upper limit on the effect of lensing.

\subsection{SMGs}

Next, we address the effects of lensing on the luminosity function of SMGs in the redshift range $z = 1-4$.  Wide field surveys conducted by the {\it Herschel} Space Observatory (Pilbratt et al. 2010) and the South Pole Telescope (Carlstrom et al. 2011) have been very effective at discovering gravitationally lensed galaxies in large numbers (e.g., Negrello et al. 2010, Vieira et al.  2013).  Lens models based on high resolution imaging with the Submillimeter Array  (SMA) (Bussmann et al. 2012, Bussmann et al. 2013) and ALMA (Hezaveh et al. 2013) are now becoming available for a substantial portion of these objects.  Therefore, it is timely to apply our approach to this population of galaxies and quantify the effect that gravitational lensing may have on its luminosity function.

\begin{figure*}
\includegraphics[width=3.4in]{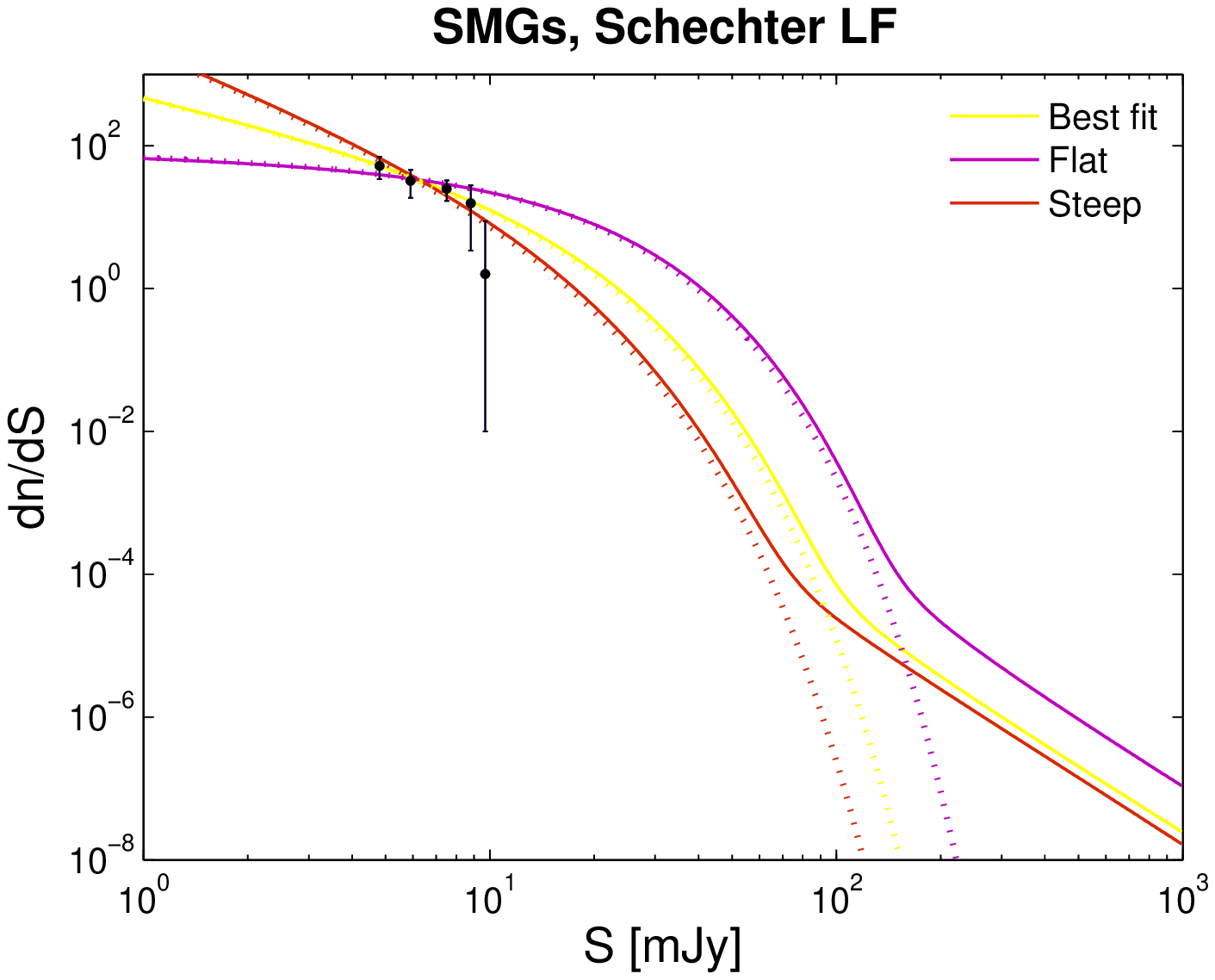}\includegraphics[width=3.4in]{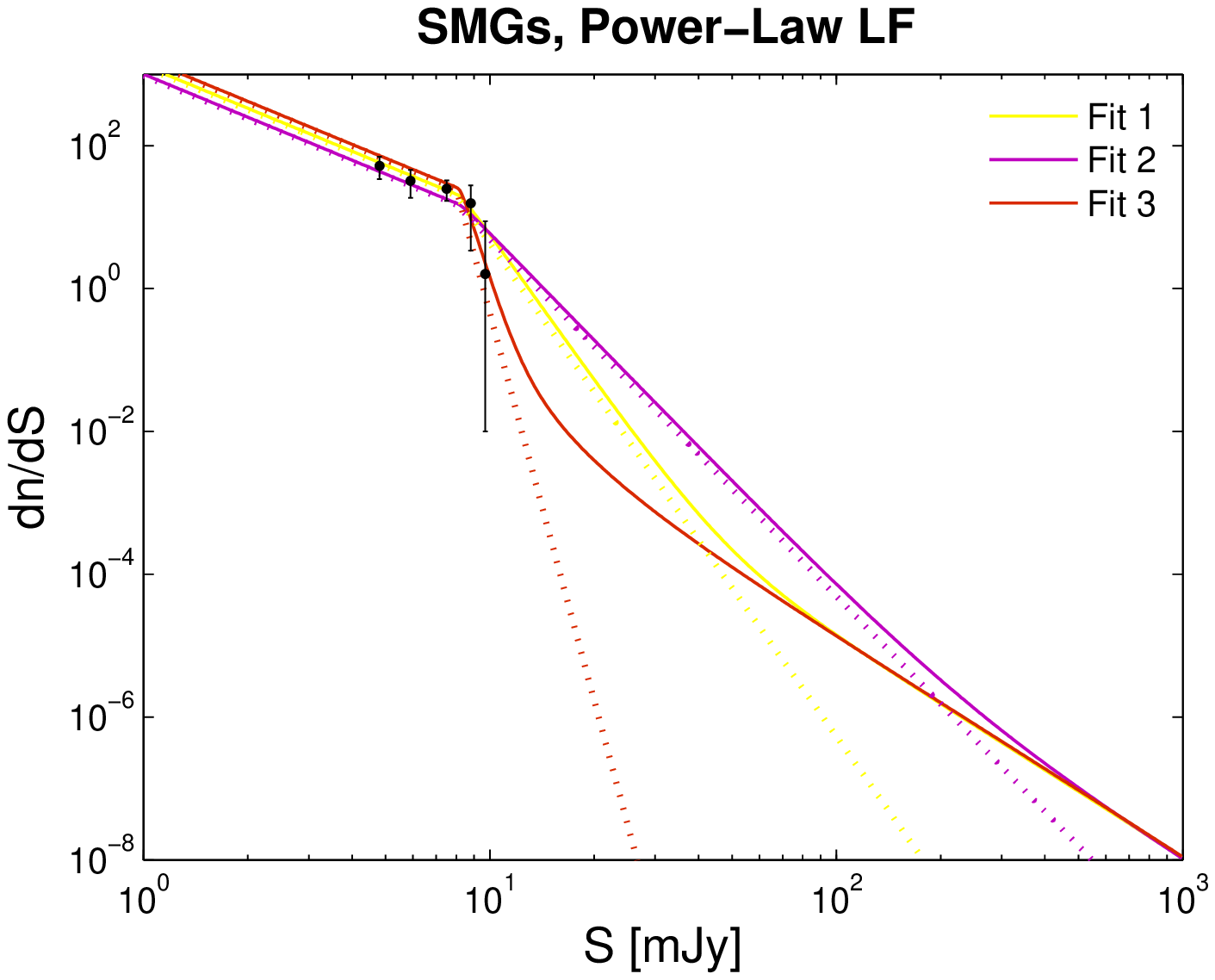}
\includegraphics[width=3.4in]{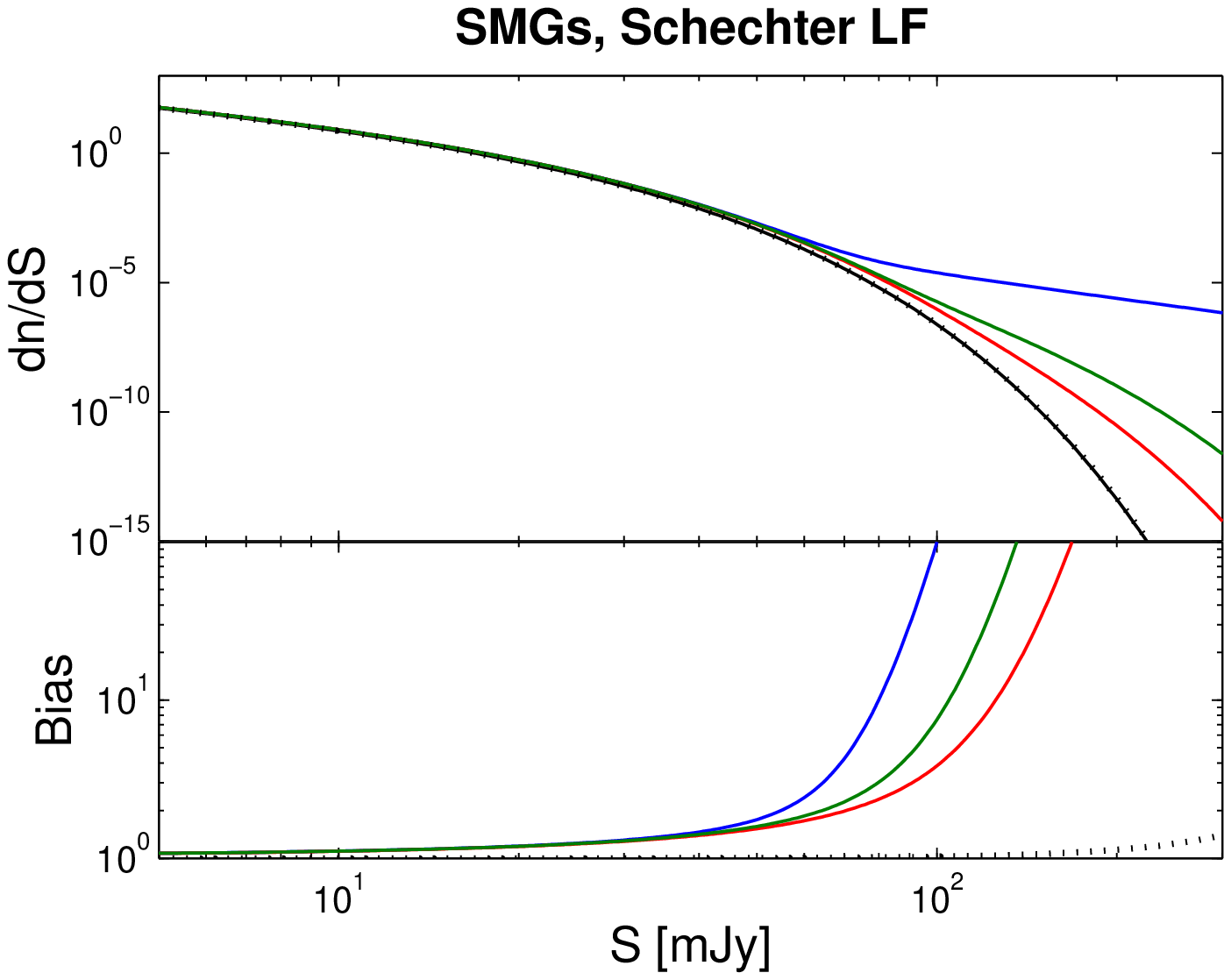}\includegraphics[width=3.4in]{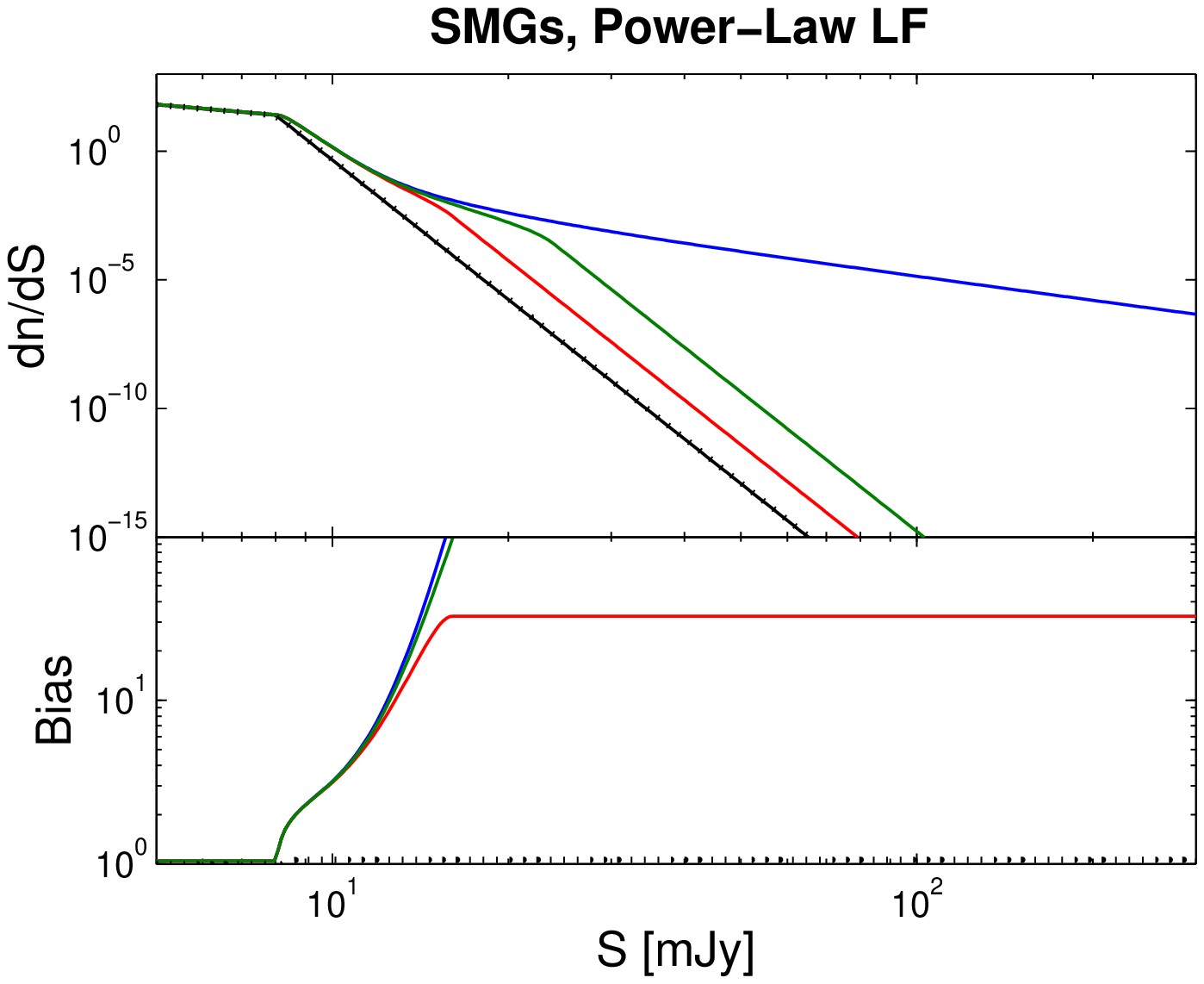}
\caption{Top row: The intrinsic (dotted) and lensed (solid, $\mu_{max} = 1000$) luminosity function of the SMGs number counts (in mJy$^{-1}$ deg$^{-2}$ units) shown for the Schechter (left) and the broken power law (right) fits  with the color code marked on each panel. The black data points are taken from the paper by \citet{Karim:2013}. Note that the number counts at the highest measured flux value, $S\sim 10$ mJy, can be underestimated due to the effect of cosmic variance. Therefore, we do not strictly account for this point when fitting the data. Bottom row: The luminosity function (top panels) and the magnification bias (bottom panels) of the SMGs fitted by our steepest fits. We show the intrinsic luminosity function (black line), and the luminosity functions lensed by  a population of virialized objects for magnifications $\mu\leq 2$ (red),  $\mu\leq 3$ (green),  $\mu\leq 1000$ (blue) as well as by proto-clusters   (black dotted). }
\label{fig:LFsm}
\end{figure*}

As mentioned above, the luminosity function of the SMGs is not very well constrained at the moment. Thus there exist a variety of functions that can equally well fit the number counts observed so far \citep{Karim:2013}.  The two most popular fits which we use here are the Schechter function and a broken power law. First, the Schechter form is generically the same one as is used in the case of the LBGs,
\begin{equation}
\frac{dn}{dS} = \frac{N^*}{S^*}\left(\frac{S}{S^*}\right)^{-\alpha}\exp\left(-S/S^*\right),
\end{equation}
with S being the observed flux, $N^*$ the normalization, $S^*$  the critical flux at which the number counts start to decline exponentially, and  $\alpha$ the faint end slope. Another frequently used functional form is the broken power law
\begin{equation}
\frac{dn}{dS} =\left\{
\begin{array}{l}
N^\star\left(\frac{S}{S^\star}\right)^{-\beta_1},~~~~\textrm{for}~~~~S<S^\star\\
N^\star\left(\frac{S}{S^\star}\right)^{-\beta_2},~~~~\textrm{for}~~~~S>S^\star
\end{array}\right.
\end{equation}
with $N^\star$ being the normalization, $S^\star$ the characteristic flux and $\beta_1$ and $\beta_2$ the slopes of the two power-laws.  In Figure \ref{fig:LFsm} we show the number counts derived by \citet{Karim:2013} using the counts of faint sub-mm galaxies in the 870-$\mu$m band of ALMA together with several  fits to the luminosity function. In particular we show three cases for the Schechter function:
\begin{enumerate}
\item Best fit  model used by \citet{Karim:2013} with $S^* = 8$ mJy, $N^* = 424$ deg$^{-2}$ and $\alpha = 1.1$,
\item Flat fit with $S^* = 10$ mJy, $N^* = 600$ deg$^{-2}$ and $\alpha = 0.1$,
\item Steep fit with $S^* = 7$ mJy, $N^* = 600$ deg$^{-2}$ and $\alpha = 1.9$,
\end{enumerate}
and three cases for the broken  power law fit:
 \begin{enumerate}
 \item Fit 1, $N^\star = 15$ mJy$^{-1}$ deg$^{-2}$, $S^\star = 8$ mJy, and $\beta_1 = 2$ and $\beta_2= 5$,
 \item  Fit 2, $N^\star = 20$ mJy$^{-1}$ deg$^{-2}$, $S^\star = 8$ mJy, and $\beta_1 =2$ and $\beta_2= 6.9$,
 \item  Fit 3, $N^\star = 25$ mJy$^{-1}$ deg$^{-2}$, $S^\star = 8$ mJy, and $\beta_1 = 2$ and $\beta_2= 18$. 
 \end{enumerate}
 As we see form Figure \ref{fig:LFsm} where we plot the intrinsic and the lensed with $\mu_{max} = 1000$ luminosity functions of SMGs together with the measured number counts \citep{Karim:2013},    the effect of lensing starts to manifest itself at $S = S^*$ in the case of the broken power law fits, while the Schechter fits do not show deviation  until $S\sim 10 S^\star$. The strength of the effect introduced by lensing depends on the steepness of the luminosity function at its bright-end. For instance, in the case of the power-law dependence, for our steepest broken power-law fit (Fit 3) the bias reaches the  value of 10 at $S \sim 2 S^\star$, while for the flattest Fit 1 it reaches same value only at $S \sim 60 S^\star$ . 
 
\begin{deluxetable*}{rcccccccc}
\tabletypesize{\scriptsize} \tablecaption{Summary of the errors introduced by gravitational lensing in the luminosity function of SMGs at $\bar z_s = 2.6$. Each entry contains the values of (Bias measured at $S = 25 S^*$,  $\Delta   N^*\%$, $\Delta S^* \%$)  introduced by lensing due to virialized halos with $\mu_{max} =$ 1.3,  1.7, 2 and 3, as well as proto-clusters, in the parameters of  the Schechter luminosity function for the intrinsic functions shown in Figure \ref{fig:LFsm}. The fits are done for the luminosities in the range $3$ mJy $<S<25 S^*$. } 
  \tablewidth{0pt}
\tablehead{ \colhead{Model} & \colhead{$\mu_{max} =    1.3$} & \colhead{$\mu_{max} =
    1.7$}&\colhead{ $\mu_{max} =2$} & \colhead{$\mu_{max} =3$} & \colhead{$\mu_{all}^{ta}$}}
\startdata 
 Flat & 4.9, 2.7$\%$, 5.9$\%$ &  20.9, 8.2$\%$, 9.9$\%$ & 56.4, 11.9$\%$, 12.7$\%$ & 511.9, 20.2$\%$, 20.0 $\%$ & 1.1, 0.5$\%$, 0.2$\%$\\ 
\hline 
Best fit & 5.5, 5.2$\%$, 6.3$\%$ & 30.4,  15.8$\%$,  11.3$\%$ & 96.6, 23.1$\%$, 15.0$\%$ & $>10^3$, 38.5$\%$, 25.6$\%$ & 1.1, 0.7$\%$, 0.2$\%$ \\
 \hline 
Steep & 6.1, 7.5$\%$, 6.6$\%$ & 41.5, 23.0$\%$, 12.5$\%$ & 151.3, 33.2$\%$, 17.1$\%$ & $>10^3$, 53.3$\%$, 30.7$\%$, & 1.1, 0.9$\%$, 0.3$\%$\\
\enddata
\label{Tab:2}
\end{deluxetable*}

 In Figure \ref{fig:LFsm} (bottom row) we compare the intermediate and strong lensing of SMGs for the steepest models in which the effect of lensing is the strongest (i.e., our steep fit in the case of the Schechter function and Fit 3 for the broken power law), showing the impact of lensing with $\mu_{max}$ = 2, 3, and 1000 on the number counts of SMGs in the case when lenses are the virialized halos as well as the case of lensing by proto-clusters. We also show the bias,  defined as the ratio of the observed to the intrinsic  number counts, $\left(dn/dS\right)_{obs}/\left(dn/dS\right)_{int}$. In addition, the effect on the derived parameters of the SMG Schechter luminosity function  and bias are further elaborated  in Table  \ref{Tab:2}  for the cases of lensing by halos with $\mu_{max} =$ 1.3, 1.5, 1.7, 2 and 3 as well as lensing by proto-clusters. As in the case of LBGs, the effect on the parameters of the observed luminosity function  depends on the upper limit of the luminosity that we expect to measure. Here we adopt the upper value of $S_{max} = 25 S^*$. Both the Figure and the Table show that in the case of SMGs at $\bar z_s =2.6$, lensing by proto-clusters has a negligible effect due to the relatively low redshift of the sources.  In particular, the bias in the case of the sources with a power-law form of the intrinsic luminosity function lensed by proto-clusters (bottom right panel of Figure \ref{fig:LFsm}) reaches 1.04 at the bright-end, while in the case of  viriallized  halos with $\mu_{max}=2$ it is $\sim 33$.  By comparing the numbers in Table \ref{Tab:2}, we find that the effect of proto-clusters in this case is negligible even when compared to lensing by halos with $\mu_{max} = 1.3$, unlike in the case of the LBGs.

When the SMGs intrinsic luminosity function is given by the broken power law, its shape is completely distorted even after it goes through a relatively mild lensig ($\mu\leq3$), as we see in Figure \ref{fig:LFsm} where the effect on Fit 3 is shown. (However, the intermediate lensing has almost no effect when a flatter power law, e.g., our Fit 1 or Fit 2 with $\beta_2 = 5$ and $6.9$, respectively, is considered). In this case a different type  of dependence (e.g., a three power-law piece-wise function) would be  needed to fit the data.   More specifically, for lensing with some $\mu_{max}$,  a second ``knee'' develops around $S_{obs} \sim \mu_{max}\times S^\star$. At the flux values  $S^{\star}<S<\mu_{max}\times S^{\star}$ the luminosity function tends to develop slope equivalent to the one in the case of strong lensing (which is $-2$ in the  case when number counts per unit flux are concerned), while at fluxes higher than $\mu_{max}\times S^{\star}$ the luminosity function retains its intrinsic power-law dependence. Thus, the effect of lensing can be easily corrected for in this case.


\section{Conclusions}

We considered the effects of gravitational lensing on the luminosity functions of Lyman-break galaxies at $z\sim 6-10$ and sub-mm galaxies at $z\sim 2-4$ in the regime when a single lens has a  dominant contribution. In particular, we made predictions for the galaxies ``in the field'', i.e., in surveyed volumes which are not strongly lensed, focusing on the intermediate range of magnifications ($\mu\leq 2$) produced by two types of lenses, namely virialized halos and non-virialized proto-clusters,  and comparing the results to the cases which do include strong lensing. The regime of intermediate lensing was not studied in the literature  and may appear to be important for future precision analysis of these two populations of sources when observed by instruments such as  HST, JWST, ALMA and {\it Herschel}.

We find that when a population of Lyman-break galaxies with a Schechter intrinsic luminosity function is subject to lensing with  magnifications $\mu\leq 3$, the observed luminosity function can still be fitted by the Schechter form, although with modified Schechter parameters. Therefore,  errors in derived intrinsic parameters are expected if de-lensing of the field galaxies is not done carefully enough. It appears however, that the common effect of lensing with  intermediate magnifications is  negligible when existing surveys with the field of view of HST are considered because of the significant cosmic variance  ($\sim 10\%$ at  $z\sim 6$  and $\sim 20\%$  at $z\sim 10$). In particular, we found that the overall effect of lensing (including the strong lensing effects) on the current Schechter parameters  \citep{Bouwens:2014} is well below 1$\%$ when only sources with  UV magnitudes above $M_{UV}\sim -21.5$ are present (as is expected for the most luminous objects in the HST surveys at $z\sim 9-12$ \citep{Oesch:2013}). However, the effect of lensing starts to be increasingly important if intrinsically brighter sources are found in the sample, which may be the case in the future with wider surveyed areas.  In this regime, the errors introduced by the weak and intermediate gravitational lensing may become significant, e.g.,  lensing with maximal magnification of 2 of a population of bright sources at $z = 12$ may result in more than $ 20\%$ discrepancy (and more than $40\%$ discrepancy when the magnifications in the range $\mu\la 3$ are considered) in the normalization of the luminosity function if sources with $M_{UV}< -24.5$ are present (while the effect on critical luminosity remains at a percent level and we keep the faint-end slope fixed). If indeed intrinsically  bright sources with $M_{UV}<-24.5$ exist so early on in the cosmic history, they would be valuable probes of the early Universe. These rare objects are expected to form in highly overdense regions, to produce stars very early, and to start heating and ionizing gas earlier than in average over the Universe.  Consequently, during the epoch of reionization ($z\geq 7$) such regions  would  create big bubbles of ionized gas over the otherwise neutral background, thus imprinting strong signature in the redshifted 21-cm signal of neutral hydrogen  which can be probed through tomography by radio telescopes such as the Square Kilometer Array \citep{LF, Ahn:2014}.

 An important point that we revealed in this paper is that particular care should be taken in the case in which   Lyman-break galaxies are lensed by massive halos that themselves host bright galaxies. In this case it is crucial to account for the magnification of the surface brightness of the source and compare it to that of the lens when deciding which sources are, in fact, observable and computing the lensing statistics. We showed that the range of impact parameters (and thus magnifications) for which the high-redshift sources are observable is reduced when the constraints from surface brightness profiles are accounted for. This leads to a suppressed probability for obtaining intermediate and strong magnifications which could result in up to a two orders of magnitude deficit in the luminosity function at high redshifts and high luminosities.

Comparing  lensing by virialized halos, which are strongly non-linear systems, to the role of proto-clusters, which we assumed to be overdense regions at the point of turnaround, we found that for a population of  Lyman-break galaxies at $z = 6-15$ the effect of proto-clusters is similar to that of lensing by massive halos with the cutoff $1.3 < \mu_{max} <1.5$  in magnification. For example,  proto-clusters can introduce $\sim 5\%$ discrepancy in the normalization of the number counts of sources at $z=12$, while lensing by virialized halos with $\mu_{max} = 1.3$ contributes only $\sim 3\%$ when the suppression in  lensing probability due to the surface brightness is taken into account (and $\sim 4\%$ when it is ignored), and lensing with $\mu_{max} = 1.5$ contributes $\sim 6\%$ ($\sim 10\%$). 

The effect of lensing on the luminosity function of sub-mm galaxies at $z\sim 2.6$ was also considered in this work. Since the intrinsic luminosity function of  this population is highly unconstrained at the moment,  the effect of lensing could appear to be either significant or negligible once this population is better explored, depending on the steepness of the luminosity function at the bright-end. In this work we considered two forms of the intrinsic luminosity function for this population, i.e., the Schechter fit and the broken power law dependence. In the latter case, lensing with intermediate magnifications may cause significant distortions of the shape of the luminosity function, and thus the effect of lensing can be easily corrected for. The former case is qualitatively very similar to that of Lyman-break galaxies, with the sole difference that there is no reduction in the lensing probability due to the large surface brightness of an extended lens. This is because the foreground lenses, which are drawn from common galaxies, are not expected to be bright at sub-millimeter wavelengths.  The discrepancy in the Schechter parameters introduced by lensing by virialized halos with intermediate magnifications   ($\mu\leq 2$) is $\sim 23\%$ in the normalization and $\sim 15\%$ in the critical luminosity when the best fit to current data is considered at flux values $3-200$ mJy. Finally, we found that the effect of lensing by proto-clusters on the galaxies at $z\sim 2-4$ is negligible when compared to the lensing with intermediate magnifications by virialized halos. For example, for the best-fit intrinsic parameters that we considered above, proto-clusters introduce $\sim 1\%$ discrepancy in the normalization of the number counts, while lensing by virialized halos with $\mu_{max} = 1.3$ contributes more than $5\%$.

 Lastly, we point out that magnification can have an important effect on the determination of the luminosity distances to the high-redshift standard candles. As a result of randomly distributed foreground  structure the luminosity distance to each source is subject to an error, which, if unaccounted for, will  bias the values of the deduced cosmological parameters.

\acknowledgments

This work was supported in part by was supported by the LabEx
ENS-ICFP: ANR-10- LABX-0010/ANR-10-IDEX- 0001-02 PSL (for A.F.) and
NSF grant AST-1312034 (for A.L.). We thank S. Bussmann and I. Dvorkin for useful discussions.


\begin{thebibliography}{}




\bibitem[Ade et al. (2014)]{Ade:2014}  Planck Collaboration; Ade, P. A. R., Aghanim, N., Armitage-Caplan, C., Arnaud, M., Ashdown, M., et al., 2014, \aap, 571, 16 

\bibitem[Ahn et al. (2014)]{Ahn:2014} Ahn, K., Xu, H., Norman, M. L., Alvarez, M., A., Wise, J., H., 2014, arXiv:1405.2085 
 
\bibitem[Alavi et al. (2014)]{Alavi:2014} Alavi, A., Siana, B., Richard, J., Stark, D. P., Scarlata, C., et al. 2014, \apj, 780, 143
 
 \bibitem[Aretxaga et al. (2011)]{Aretxaga:2011}  Aretxaga, I., Wilson, G. W., Aguilar, E., Alberts, S., Scott, K. S.,
 2011, \mnras, 415, 3831
 
 \bibitem[Atek et al. (2015)]{Atek:2015}  Atek, H., Richard, J., Kneib, J.-P., Jauzac, M., Schaerer, D., 2015, \apj, 800, 18

 
\bibitem[Barkana \& Loeb (2000)]{Barkana:2000} Barkana, R., \&  Loeb, A., 2000, \apj, 531, 613
 \bibitem[Barkana \& Loeb (2001)]{Barkana:2001} Barkana, R., \& Loeb, A., 2001, PhR, 349, 125

\bibitem[Barone-Nugent et al. (2014)]{Barone:2014} Barone-Nugent, R. L., Trenti, M., Wyithe, J. S. B., Bouwens, R. J., Oesch, P. A., et al., 2014, \apj, 793, 17


 \bibitem[Behroozi, Wechsler \& Conroy (2013)]{Behroozi:2013} Behroozi, P., S., Wechsler, R. H.,  \& Conroy, C., 2013, \apj, 770, 57

 \bibitem[Behroozi \& Silk (2015)]{Behroozi:2014} Behroozi, P., S. \& Silk, J., 2015, \apj, 799, 32 

 \bibitem[Blandford \& Narayan (1992)]{Blandford:1992} Blandford, R., D. \& Narayan, R., 1992, ARA$\&$A, 30, 311

 \bibitem[Blain et al. (1999)]{Blain:1999} Blain A. W., Kneib J.-P., Ivison R. J., Smail I., 1999, \apj, 512, L87

  \bibitem[Bouwens et al.(2009)]{Bouwens:2009} Bouwens, R. J., Illingworth, G. D., Bradley, L. D., Ford, H., Franx, M.,  et al., 2009, \apj, 690, 1764.


 \bibitem[Bouwens et al.(2014)]{Bouwens:2014} Bouwens, R. J.,  Illingworth, G. D.,
  Oesch, P. A., Trenti, M., Labbe, I., et al.,  2014, arXiv:1403.4295
  
   
     \bibitem[Bruzual \& Charlot (2003)]{Bruzual:2003} Bruzual, G. \& Charlot, S.,  2003, \mnras,
344, 1000

\bibitem[Bussmann et al. (2012)]{Bussmann:2012} Bussmann, R. S., Gurwell, M. A., Fu, Hai, Smith, D. J. B., Dye, S., et al.,  2012, \apj, 756, 134

\bibitem[Bussmann et al. (2013)]{Bussmann:2013} Bussmann, R. S., Perez-Fournon, I., Amber, S., Calanog, J., Gurwell, M. A., et al.,  2013, \apj, 779, 25

\bibitem[Bussmann et al. (2015)]{Bussmann:2015} Bussmann, R. S.,  et al.,  2015, in preparation.


\bibitem[Carlstrom et al. (2011)]{Carlstrom:2011} Carlstrom, J. E., Ade, P. A. R., Aird, K. A., Benson, B. A., Bleem, L. E.,  2011, PASP, 123, 568

\bibitem[Chapman et al. (2005)]{Chapman:2005} Chapman S. C., Blain A. W., Smail I., Ivison R. J., 2005, \apj, 622, 772

 \bibitem[Conroy et al.(2006)]{Conroy:2006} Conroy, C., Wechsler, R. H., \& Kravtsov, A. V. 2006, \apj,
647, 201
    
\bibitem[Comerford et al.(2002)]{Comerford:2002} Comerford, J. M., Haiman, Z., \& Schaye, J., 2002, \apj, 580, 63

\bibitem[Ellis(2014)]{Ellis} Ellis, R., Proc. of the 26th Solvay
  Conference in Physics, R. Blandford and A. Sevrin, eds., World
  Scientific; arXiv:1411.3330

 \bibitem[Fazio et al. (2004)]{Fazio:2004} Fazio, G. G., Hora, J. L., Allen, L. E., Ashby, M. L. N., Barmby, P., et al., 2004, ApJS, 154, 10

 \bibitem[Gonzalez-Nuevo et al. (2012)]{Gonzalez-Nuevo:2012} Gonzalez-Nuevo, J., Lapi, A., Fleuren, S., Bressan, S., Danese, L., 2012, \apj, 749, 65

\bibitem[Hezaveh et al. (2013)]{Hezaveh:2013} Hezaveh, Y. D., Marrone, D. P., Fassnacht, C. D., Spilker, J. S., Vieira, J. D., 2011 \apj, 767, 132

\bibitem[Hughes et al. (1998)]{Hughes:1998} Hughes, D. H., Serjeant, S., Dunlop, J., Rowan-Robinson, M.,
 Blain, A., et al., 1998, \nat, 394, 241


 \bibitem[Jain \& Lima (2011)]{Jain:2011} Jain, B. \& Lima, M., 2011 \mnras, 411, 2113
 
 \bibitem[Karim et al. (2013)]{Karim:2013} Karim, A., Swinbank, A. M., Hodge, J. A., Smail, I. R., Walter, F., et al., 2013, \mnras, 432, 2


\bibitem[Kimble et al. (2008)]{Kimble:2008} Kimble, R. A., MacKenty, J. W., O’Connell, R. W., \& Townsend,
J. A. 2008, Proc. SPIE, 7010, 70101E


\bibitem[Kochanek  (1994)]{Kochanek:1994} Kochanek, C. S., 1994, \apj, 436, 56



\bibitem[Kravtsov et al.(1998)]{Kravtsov:1998} 
Kravtsov, A. V., Klypin, A. A., Bullock, J. S., Primack, J. R. 1998, ApJ,
502, 48.

  \bibitem[Kravtsov et al. (2004)]{Kravtsov:2004}  Kravtsov, A. V., Berlind, A. A., Wechsler, R. H., Klypin,
A. A., Gottlober, S., et al.,  2004,
\apj, 609, 35

  \bibitem[Kravtsov (2013)]{Kravtsov:2013} Kravtsov, A. V.,  2013, \apj, 7644, 31.

  \bibitem[Kurczynski et al.(2014)]{Kurczynski:2014} Kurczynski, P., Gawiser, E., Rafelski, M., Teplitz, H. I., Acquaviva, V., et al., 2014, \apj, 793, 5

 \bibitem[Lapi et al.  (2011)]{Lapi:2011} Lapi, A., Gonzalez-Nuevo, J., Fan, L., Bressan, A., De Zotti, G., 2011 \apj 742 24

 \bibitem[Lima et al.  (2010a)]{Lima:2010a} Lima, M., Jain, B., Devlin, M. \& Aguirre, J., 2010, \apj, 717, 31

 \bibitem[Lima, Jain \& Devlin (2010b)]{Lima:2010b} Lima, M., Jain, B. \& Devlin, M., 2010, \mnras, 406, 2352

\bibitem[Loeb \& Furlanetto (2013)]{LF} Loeb, A. \& Furlanetto, S.,
  2013, The First Galaxies in the Universe, Princeton University Press   (Princeton)
 
 \bibitem[Madau et al. (1996)]{Madau:1996}Madau, P., Ferguson, H. C., Dickinson, M. E., Giavalisco, M.,
Steidel, C. C., \& Fruchter, A., 1996, \mnras, 283, 1388

 \bibitem[Madau, Pozzetti \& Dickinson (1998)]{Madau:1998} Madau, P., Pozzetti, L. \& Dickinson, M., 1998, \apj, 498, 106

 \bibitem[Mandelbaum et al. (2005)]{Mandelbaum:2005} Mandelbaum, R., Tasitsiomi, A., Seljak, U., Kravtsov, A. V., Wechsler, R. H.,  2005, \mnras, 362, 1451

\bibitem[Mason et al. (2015)]{Mason:2015} Mason, C. A., Treu, T., Schmidt, K. B., Collett, T. E., Trenti, M., et al.,   2015, arXiv:1502.03795

\bibitem[McLure et al. (2013)]{McLure:2013} McLure, R. J., Dunlop, J. S., Bowler, R. A. A., Curtis-Lake, E., Schenker, M., et al. 2013, \mnras, 432, 2696

\bibitem[Moster et al.(2011)]{Moster:2011} Moster, B. P.,
    Somerville, R. S., Newman, J. A.,  \&  Rix, H.-W., 2011, \apj, 731, 113

\bibitem[Navarro, Frenk \& White (1997)]{Navarro:1997} Navarro, J. F., Frenk, C. S., \& White, S. D. M., 1997, \apj, 490, 493
\bibitem[Negrello (2007)]{Negrello:2007}  Negrello, M., Perrotta, F., Gonzalez-Nuevo, J., Silva, L., de Zotti, G., et al. 2007, \mnras, 377, 1557

\bibitem[Negrello et al. (2010)]{Negrello:2010}  Negrello, M., Hopwood, R., De Zotti, G., Cooray, A., Verma, A., et al. 2010, Science, 330, 800



\bibitem[Newman \&  Davis (2002)]{Newman:2002} Newman, J. A., \& Davis, M. 2002, \apj, 564, 567


\bibitem[Oesch et al. (2013)]{Oesch:2013}  Oesch, P. A.,  Bouwens, R. J.,  Illingworth,  G. D., Labbé, I., Franx, M., et al., \apj, 773, 75.
 \bibitem[Oesch et al.(2014)]{Oesch:2014} Oesch, P. A., Bouwens, R. J., Illingworth, G. D., Labbe, I., Smit, R., et al. 2014, \apj, 786, 108.
    \bibitem[Oteo et al.(2014)]{Oteo:2014}   Oteo, I., Bongiovanni, A., Magdis, G., Perez-Garcia, A. M., Cepa, J., et al., 2014, \mnras, 439, 1337 

\bibitem[Pei (1993)]{Pei:1993}  Pei, Y. C.,   \apj, 404, 436.

\bibitem[Pei (1995)]{Pei:1995}  Pei, Y. C.,   \apj, 440, 485.


\bibitem[Perrotta  et al. (2002)]{Perrotta:2002} Perrotta, F., Baccigalupi, C., Bartelmann, M., De Zotti, G., \& Granato, G. L. 2002, \mnras, 329, 445

\bibitem[Pilbratt et al. (2010)]{Pilbratt:2010} Pilbratt, G. L., Riedinger, J. R., Passvogel, T., Crone, G., Doyle, D., G. L. 2002, A \& A, 518, L1



\bibitem[Schechter (1976)]{Schechter:1976} Schechter, P. 1976, \apj, 203, 297

\bibitem[Schenker et al. (2013)]{Schenker:2013} 	Schenker, M. A., Robertson, B. E., Ellis, R. S., Ono, Y., McLure, R. J., et al. 2013, ApJ, 768, 196

\bibitem[Sheth \& Tormen (1999)]{Sheth:1999} Sheth, R. K., \& Tormen, G., 1999, \mnras, 308, 119

\bibitem[Simpson et al. (2014)]{Simpson:2014} Simpson, J. M., Swinbank, A. M., Smail, I., Alexander, D. M.,
Brandt, W. N., et al.,  2014, \apj, 788, 125

\bibitem[Somerville et al. (2004)]{Somerville:2004} Somerville, R. S., Lee, K., Ferguson, H. C., Gardner, J. P., Moustakas, L. A., \& Giavalisco, M. 2004, \apj, 600, L171


 \bibitem[Szomoru et al.  (2013)]{Szomoru:2013} Szomoru, D., Franx, M., van Dokkum, P. G., Trenti, M., Illingworth, G. D., et al., 2013, \apj, 763, 73.
 
  \bibitem[Tasitsiomi et al. (2004)]{Tasitsiomi:2004}  Tasitsiomi, A., Kravtsov, A. V., Wechsler, R. H., \& Primack, J. R. 2004, \apj, 614, 533.


  \bibitem[Trenti \& Stiavelli (2008)]{Trenti:2008} Trenti, M., Stiavelli, M.  2008, \apj, 676, 767.
  
\bibitem[Turner et al. (1984)]{Turner:1984}  Turner, E. L., Ostriker, J. P.,  \& Gott III, J. R., 1984, \apj, 284, 1.

  
  
 
 \bibitem[Vale \& Ostriker (2004)]{Vale:2004} Vale, A., \& Ostriker, J. P., 2004, \mnras, 353, 189

 \bibitem[Vieira et. al. (2013)]{Vieira:2013} Vieira, J. D., Marrone, D. P., Chapman, S. C., De Breuck, C., Hezaveh, Y. D., et al.,  2013 \nat 495, 344

 \bibitem[Wardlow et al.(2013)]{Wardlow:2013} 	Wardlow, J., L., Cooray, A., De Bernardis, F., Amblard, A., Arumugam, V., 2013 \apj 762, 59 
 
   \bibitem[Webster et al.(1988)]{Webster:1988} Webster, R. L., Hewett, P. C., Harding, M. E., and Wegner, G. A.,  1988, \nat, 336, 358.

 \bibitem[Wyithe et al.(2011)]{Wyithe:2011} Wyithe, J. S. B.,
 Yan, H., Windhorst, R. A., \& Mao, S., 2011, \nat, 496, 7329, 181
 
 \end{thebibliography}
\end{document}